\newcommand{\tr}[1]{\hbox{ {\raisebox{0.0cm}{Tr}
{\tiny\raisebox{-0.25cm}{$\!\!\!\!\!\!\!\!\!\!\!\!\!\!\: \{#1\}$}} 
$\!\!\!\!\!\!$
} } }
\newcommand{\gp}{\,.}
\newcommand{\gk}{\,,}
\newcommand{\fxi}{\mbox{\boldmath $\xi$}}
\newcommand{\fw}{\mbox{\boldmath $w$}}
\newcommand{\beqn}{\begin{equation}}
\newcommand{\eeqn}{\end{equation}}
\newcommand{\beqa}{\begin{eqnarray}}
\newcommand{\eeqa}{\end{eqnarray}}

\newcommand{\xib}{\mbox{\boldmath $\xi$}}
\newcommand{\wb}{\mbox{\boldmath $w$}}

\newcommand{\lsim}{\hbox{ {\raisebox{0.06cm}{$<$}
\raisebox{-0.14cm}{$\!\!\!\!\!\!\!\!\: \sim$}} } }

\documentstyle[11pt]{article}
\let\al =\alpha
\let\bet=\beta
\let\gam=\gamma
\let\sig=\sigma
\let\delt=\delta

\let\eps=\epsilon

\title{Phase transitions in the generalization behaviour of
 multilayer perceptrons:\\II. The influence of noise
\thanks{based on the PhD thesis of B.~Schottky,
 Regensburg 1996}}
 \author{B Schottky$^1$ and U Krey$^2$
\\  \\$^1$ Department of Comp.\ Science and Appl.\ Math., 
Aston University,  
\\Birmingham B4 7ET, UK 
\\ \\$^2$ Institut f\"ur Physik II der Universit\"at Regensburg,\\
Universit\"atsstr.~31, D-93040 F.R.G. }
       
\begin{document}

\large
\date{October 3, 1997, to appear in J.~Phys.~A} 
\maketitle
 
\begin{abstract} We extend our study of phase transitions in the
generalization behaviour of multilayer perceptrons with non-overlapping
receptive fields to the problem of the {\it influence of noise}
concerning e.g.\ the input units and/or the couplings between the input
units and the hidden units of the second layer (='input noise') or the
final output unit (='output noise'). Without output noise, the output
itself is given by a general, permutation-invariant Boolean
function of the outputs of the hidden units.  As a  result we find that
the phase transitions which we found in the deterministic case, mostly
{\it persist} in the presence of noise. The influence of the noise on
the position of the phase transition, as well as on the behaviour in
other regimes of the loading parameter $\alpha$, can often be described
by a simple rescaling of $\alpha$ depending on strength and type of the
noise. We then consider the problem of the optimal noise level for
Gibbsian and Bayesian learning, looking on replica symmetry breaking as
well. Finally we consider the question why learning with errors is
useful at all.

\end{abstract}

\section{Introduction and overview}
\subsection{Introduction and basic definitions}
In a recent paper, \cite{l:scho1}, one of us has treated the problem of
phase transitions in the generalization behaviour of two-layer neural
networks with non-overlapping receptive fields. The architecture of the
systems considered is shown in Fig.\ 1. It corresponds to a tree of totally
$N$ input units, which are grouped into $K$ vectors
$\xib_1$,..., $\xib_K$ of $M:=N/K$ binary components
$\xi_{k,m}=\pm 1$, with $k=1,...,K$ and $m=1,...,M$.

Each one of these vectors $\xib_k$ determines the binary output
$\sigma_k$ of a so-called ''hidden unit'' according to the 
perceptron-rule
\beqn
\sigma_k = {\rm{sgn}} \left ( \frac{1}{\sqrt{M}}
\wb_k\cdot \xib_k \right ) \equiv
{\rm{sgn}}
\left (\frac{1}{\sqrt{M}}\sum_{m=1}^M w_{k,m}\cdot \xi_{k,m} \right )\,\,,
\eeqn
where the so-called coupling vectors $\wb_k$ have $M$ arbitrary real
components $w_{k,m}$, which are only constrained by the normalization
$\wb_k^2=M$.

The final output $\sigma$ (='classification', 'answer') of the machine
 for a given 
input (='question') results from a fixed Boolean function
\beqn\label{eq2}
\sigma =B(\sigma_1,...,\sigma_k)\equiv B(\{\sigma_k\})
\eeqn
of the outputs $\sigma_k$ of the hidden units.
This Boolean function is
 arbitrary apart from the postulate that it should be invariant against
a permutation of the arguments. 

Now the task of this classification machine is to learn a certain
''rule'' by {\it modification of the coupling vectors} through learning
the correct classification of a set of input examples $\xib^\mu
=(\xib_1^\mu,...,\xib_K^\mu)$, with $\mu=1,...,p$. Here it is assumed
that the so-called loading parameter $\alpha :=p/N$ is finite, while the
thermodynamic limit $N\to\infty$ is taken.
  
In the following it is also assumed that
the ''rule'', by which the correct answers follow from the
questions, is implemented by a ''teacher perceptron'' of the same
architecture as given above, with fixed ''teacher couplings''
$\wb^{t}$. In particular we assume that the Boolean function of the
student machine is the same as that of the teacher. However, the noise
levels can be different, unless otherwise stated (see below).

We consider the {\it generalization ability} $g(\alpha )$, see
\cite{l:watkin,l:opperkinzel}, of the system
after a training process with $p =\alpha\cdot N$ examples; $g(\al )$ is
defined as the probability that after the training an additional random
question is answered correctly, i.e.\ in the same way as the teacher
would answer in the absence of noise. It should be stressed that after
the training we switch off {\it any} noise, both for the teacher and for
the student machine. In contrast, during the training, noise of various
kind will corrupt both the student and the teacher behaviour (see
below).

Of course $g(\al )$ generally does not only depend on $\al$, but also on
the architectures considered, i.e.\ on the Boolean function
$B(\{\sigma_k\})$, and on the noise. Only in the limit $\al\to\infty$, as
already shown in \cite{l:scho1}, in the absence of noise the
architecture does not matter, and one obtains for $\al\to \infty$
asymptotically the universal result
 \beqn\label{eq3}
  g(\al )\to 1 -\frac{0.625}{\al}
  \eeqn
for the so called Gibbsian learning (see below), where a student is drawn
randomly from an ensemble which consists in the deterministic case just of
those students classifying the training set correctly.

This asymptotic result is independent of the choice of the Boolean
function and the number $K$ of hidden units. Although, as already
mentioned, our Boolean functions are quite general,
 apart from the constraint of permutation invariance, and although the
behaviour depends essentially only on a small set of characteristic
numbers (see below), we mention for the following that the most
important machines considered are

\begin{itemize}
\item the {\it committee machine}: This
machine classifies by a majority vote of the $K:=2n+1$ hidden units,
i.e.\ with

\beqn\label{eq4}
\sigma = {\rm sgn} \left ( \sum_{k=1}^K \sigma_k \right )\,\,,
\eeqn

\item whereas the {\it parity machine} is defined for general $K\ge 2$ by

\beqn \label{eq5}
\sigma = {\rm sgn} \prod_{k=1}^K \sigma_k\,\,\,,
\eeqn

\item and finally the AND-machine by
\beqn\label{eq6}
\sigma ={\rm sgn} \left ( \sum_{k=1}^K \sigma_k -K+1 \right )\,\,\,,
\eeqn
i.e.\ a positive classification is only given, if all hidden units
agree.
\end{itemize}

The main result of the present paper concerns the possible existence of
phase transitions in the generalization behaviour as a function of $\al$.
E.g.\ for the {\it parity} machine, in contrast to the {\it committee},
generalization starts only if $\al $ is larger than a critical value,
\cite{l:HanselMato,l:Schwarze,l:Opper} (''Aha
effect''). In the preceding paper, \cite{l:scho1},  this was discussed
for the deterministic case, whereas in the present second part we
discuss the influence of noise. Generally, we find that the phase
transitions mostly persist, although with changed critical values, and we also
find certain scaling laws combining the critical loading parameter $\al$
and the ''noise strength''. Furthermore, the performance of the
system is found to be optimal, when the noise of the ''student machine''
adapts to that of the teacher in a certain way.

\subsection{Overview}

Since there are a lot of categories considered in this paper, it is
easy to lose track. So we give a brief overview for better
orientation. The categories considered are:
\begin{itemize}
\item Two types of noise, {\it input-} and {\it output-}noise, see
subsections 2.5 and 2.6 below.
\item We consider (i) the case that noise-levels of teacher and student are
assumed to be {\it the same} (chapter 3), but also the case
(ii) that the student noise
level can be {\it chosen to optimize} the learning behaviour (chapter 4).
 \item
Although most results are discussed for general values of the reduced
size $\alpha :=p/N$ of the training set, the two limits $\alpha\to 0$
(more precise would be: '$\alpha$ small', see subsection 3.2.1) 
and $\alpha\to\infty$ are of special interest. 
\item
Concerning the noise strength, particular emphasis is put on the two
limits of {\it small} and {\it large} noise level, respectively.
\item
Our main results are for general two-layer perceptrons with $K$
''hidden units'', where $K>1$; however, emphasis is sometimes put on the
case  $K=1$, i.e.~the single-layer perceptron.
\item There are different
learning rules (section 2.2 below), and for $K>1$ one has
to distinguish the different Boolean output functions (section 2.3).
\item Our main results are obtained with the
{\it replica-symmetric} approach (see below), but we also discuss
some results obtained  with {\it broken
replica-symmetry} (see chapter 4).
 \end{itemize}
 Since in principle all these categories can be  arbitrarily combined
there is a very large number of combinations, but not all of them are
considered in this paper. We mention as well that some considerations
are {\it only} made for the simple perceptron ($K=1$).\newline
The paper is now organized as follows:

\begin{itemize}
\item Section 2 outlines the theoretical framework used in this paper,
describes the {\it learning algorithms} considered and introduces the
two {\it types of noise} treated in this paper.
\item The whole of section 3 deals with the case (i) mentioned above,
i.e.~there it is assumed that the {\it noise level}
 for the student is chosen
to be {\it the same} as that of the teacher (which is not a bad choice).
Both types of noise are considered, with special emphasis on the
limiting cases $\alpha\to0$ and $\alpha\to\infty$ of the loading
parameter.
\item In section 4 we investigate the impact of a {\it varying} student
noise level at fixed teacher noise, i.e.~case (ii), aiming at the
optimal choice. This is done only for the case of output noise.
Furthermore, here we concentrate mainly on the single-layer perceptron,
taking  'replica symmetry breaking' into account. The multilayer case is
treated only for the limiting cases of large and small training sets.
\item Section 5 deals with the question, why a finite student noise
level (which means that some of the training patterns are not learned
correctly) can be useful at all.
\item Finally section 6 presents our conclusions. \end{itemize}

\section{Basic theory}

The answering behaviour of a student and a teacher network
for given weights is defined by the two functions
\beqn
\phi_{s/t}(\sigma^\mu|\wb^{s/t},\xib^\mu )
\eeqn
determining the probability of getting the final answer $\sigma^\mu$ on a
question $\xib^\mu$ if the weights $\wb$ are given.
The sub-/superscripts are 's' and 't' for student and teacher, respectively. 
So $\phi$ encodes both the underlying architecture and the noise process
corrupting the answer.

\subsection{The version space and free energy}

From $\phi_t$ one derives the probability $P_t$ of getting answers  
$\{\sigma^\mu\}$ for patterns $\{\xib^\mu\}$, with $\mu=1,...,p$,
 by the teacher rule:
\beqa
P_t(\tilde\sigma^p)&=&\int {\rm d}\fw^tP_w(\fw^t)\cdot p(\tilde
\sigma^p|\fw^t)\nonumber\\
&=&\int{\rm d}\fw^tP_w(\fw^t)\prod_{\mu=1}^p\phi_t(\sigma^\mu|\fw^t,\fxi^\mu)\gk
\eeqa
where the so-called {\it prior} $P_w$ takes care of the normalization
constraints. Here and in the following we use $\tilde\sigma^p$ as
notion for the set $\{\sigma^\mu\}$ of answers, with $\mu =1,...,p$.

Using the Bayes theorem, the probability that a specific
 weight vector $\fw^s$
is the correct one given the training patterns and the answers by the
teacher, is determined through $\phi_s$ by
\beqn
p(\fw^s|\tilde\sigma^p)=\frac{P_w(\fw^s)p(\tilde\sigma^p|\fw^s)}
{\int {\rm d}\fw^sP_w(\fw^s)p(\tilde\sigma^p|\fw^s)}\gp
\label{wfact}
\eeqn
with
\beqn
p(\tilde
\sigma^p|\fw^s)=\prod_{\mu=1}^p\phi_s(\sigma^\mu|\fw^s,\fxi^\mu)\gp
\eeqn
This defines the {\it degree of membership} of a specific coupling vector
 $\fw^s$ to the so called {\it version space}.
 The version space contains all student coupling
vectors with a weight proportional to the probability that these
 couplings agree with those of the actual teacher.
  So one defines the corresponding partition function
\beqn
Z(\tilde\sigma^p)=\int {\rm d}\fw^s P_w(\fw^s)
\prod_{\mu=1}^p\Phi_s(\fw^s|\sigma^\mu,\fxi^\mu)\gk
\eeqn
where
\beqn
\Phi_s(\fw^s|\sigma,\fxi)=c_E\phi_s(\sigma|\fw^s,\fxi)\gk
\eeqn
with $c_E$ being a positive free constant, which takes into account that the
 degree of membership has not to be normalized.

In our case we restrict
the possible functions $\phi_{s/t}$ of the student resp.\ the
teacher to depend only on
the corresponding {\it local field values}
 $h^{s/t}_k$ of the hidden nodes,
\beqn
h^{s/t}_k :=\frac{1}{\sqrt{M}}\fw^{s/t}_k\cdot\fxi_k\gk
\eeqn
so
\beqn
\phi_{s/t}(\sigma|\fw^{s/t},\fxi)
\equiv\phi_{s/t}(\sigma|\{h^{s/t}_k\})\gp
\eeqn
We can now perform the Gardner analysis, \cite{l:gardner},
 by calculating
\beqa
{\cal F} :&=&\frac{1}{N}\langle\langle\ln Z\rangle_{\fw^t}
\rangle_{\{\fxi^\mu\}_{\mu=1\ldots p}}
\nonumber\\ \label{freieansatz}
&=&\frac{1}{N}\tr{\tilde\sigma^p}\langle P_t(\tilde\sigma^p)\ln
Z(\tilde\sigma^p)
\rangle_{\{\fxi^\mu\}_{\mu=1\ldots p}}\,,
\eeqa
which we will call 'free energy' although this is physically not precise.

 To describe the structure of the version space in the thermodynamic
limit $N\to\infty$ we introduce the order parameters
\beqa
q_k&:=&\frac{1}{M}\fw^1_k\cdot\fw^2_k\\
r_k&:=&\frac{1}{M}\fw^t_k\cdot\fw^s_k\gp
\eeqa
So $q_k$ is the overlap between the k-th subperceptron of two students
chosen
randomly from the version space, and $r_k$ is the corresponding overlap of 
a random student vector with the teacher couplings. Nevertheless, since
$B$ is restricted to be permutation symmetric, these quantities
cannot depend
on the node number $k$, and thus it is sufficient to use
\beqn \label{eqn20}
q=q_k\quad ;\quad r=r_k
\eeqn
as permutation-symmetric order parameters.
In the replica-symmetric approximation, straightforward calculations,
see \cite{l:schottkydiss}, lead to the final result
\beqn               \label{eq27}
{\cal F}={\rm extr}_{(q,r)}\left \{
K^{-1}\sum_{k=1}^K \left [ \frac{1}{2} \ln (1-q)
+\frac{1}{2}\frac{q-r^2}{1-q} \right ]
-\al\,W(q,r)
\right \}\,,
\eeqn
with the so-called {\it energy term}
\beqn\label{eq28}
W(q,r) =\int\prod_{k=1}^K {\rm D}t_k\sum_{\sig=\pm
1}F_t(\sig,\{q,r,t_k\})\,\cdot\ln F_s(\sig,\{q,t_k\})\,,
\eeqn
and the {\it architecture specification} for the teacher machine
\beqn            \label{eq29}
F_t(\sig,\{q,r,t_k\})=\int \prod_{k=1}^K {\rm D}s_k
\phi_t\left (\sig,\{s_k\sqrt{1-\frac{r^2}{q}}
-t_k\frac{r}{\sqrt{q}}\}\right )
\eeqn
and for the student machine
\beqn \label{eq30}
F_s(\sig,\{q,t_k\})=\int\prod_{k=1}^K {\rm D}s_k \phi_s
\left (\sig,\{s_k\sqrt{1-q}-t_k\sqrt{q}\}\right )\,.
\eeqn

Here ${\rm D}x=(2\pi)^{-1/2}\,{\rm d}x\,\exp(-x^2/2)$ is the Gauss
measure, and the values of the order parameters $q,r$ are fixed
by the saddle-point conditions
\beqn\label{eq31} 
\frac{\partial {\cal F}}{\partial q} = \frac{\partial
{\cal F}}{\partial r} =0\,.
\eeqn
These are very general formulas which allow to calculate % the  evolution of
the order parameters for classification machines with tree architecture.
Nevertheless, for {\it non}-permutation-symmetric Boolean functions $B$ one
would have to distinguish between the $q_k,r_k$ for different nodes.

\subsection{Gibbs and Bayes algorithms; Gardner analysis}

As training algorithms we discuss (i) the Gibbs algorithm and (ii) the
Bayes algorithm. They are distinguished by the way, how the version
space is utilized: For the Gibbs algorithm, a ''typical student
machine'' is drawn at random out of the version space according to the
weight factor (\ref{wfact}), and then an average is performed as usual.
In contrast, for the Bayes algorithm one takes into account {\it all}
members of the version space and gives that answer, which corresponds to
their weighted majority vote: In this way, the a-posterior error
probability is minimized. Therefore, the answers given by this so-called
Bayes procedure usually cannot be obtained from only one machine of the
kind considered.

\subsection{Notation to encode the Boolean function}

For a specific Boolean function $B$ (with a given number $K$ of hidden
units) we define the following expression~:
\beqn
\Delta^\sigma(\{\sigma_l\})=\left\{\begin{array}{r@{\quad:\quad}l}
1 & B(\{\sigma_l\})=\sigma\\
0 & \mbox{sonst}
\end{array}
\right.
\label{dtdef}
\eeqn
Thus $\Delta^\sigma$ is 1 just for those internal representations which
are mapped to '$\sigma$' by the Boolean function $B$. We remind that
only learnable problems are considered, so the same Boolean function
specifies the architecture for the teacher and student networks, and thus
we need no superscript to distinguish between them.

We need as well a short code to denote a special architecure. We use
the same convention as already introduced in \cite{l:scho1}: A Boolean
function is characterized by its number of nodes {\it K} and a special
'mcode' {\it q} with $q=\sum_{\nu=0}^{K-1}n_\nu2^\nu$. Here $n_\nu=0$ or
$=1$, respectively, if a positive vote of exactly $\nu$ hidden units
leads to a negative resp.\ positive final output $\sigma$ of the Boolean
function. (By convention, $\nu=K$ shall always imply a positive output.)
The name is then combined to 'K{\it K}\_mcode{\it q}', so for example
'K4\_mcode2' is a network with 4 hidden units and positive output if
exactly 4 or 1 hidden unit(s) have positive vote.

\subsection{Error probability}

The order parameters describe pretty well the learning success (or
failure) and we will often present just these values. Nevertheless the
quantity, which is so to say of final interest, is the generalization
ability. In the presence of noise, there exist of course different
possibilities to define this quantity; our choice is, as stated already
above, to assume that {\it after training} the noise is switched-off
completely, both for the teacher and for the student networks. Moreover,
we refer rather to the generalization {\it error} $\eps(\alpha):=1-g(\al)$
measuring the probability that student and teacher {\it disagree} on a
new question.

For a given architecture this generalization error is determined
uniquely by the values of the order parameters $q(\alpha)$ and
$r(\alpha)$, no matter which noise processes have influenced the
learning.

For the Gibbs algorithm $\eps$ depends only on the typical 
student-teacher overlap $r$ and is given by (see as well eqn.\ (21) 
in Ref.\ [1]): 
\beqn\label{eq32}
\eps(r)=\left (\frac{1}{2}\right  )^K\sum
_{\sig=\pm 1}\left [
{\rm \tr{\sig_k^t}}{\rm
\tr{\sig_k^s}}\Delta^\sig(\{\sig_k^t\})
\Delta^{-\sig}(\{\sig_k^s\})\prod_k\left (
1 - \frac 1{\pi}\arccos(\sig_k^t\sig_k^s r)
\right )
\right ] \,.
\eeqn

For the Bayes algorithm we obtain the generalization error by
\beqn\label{eq85}
\eps_{{\rm\scriptsize{Bayes}}} =\int \prod_{k=1}^K {\rm D}t_k\,\, {\rm min} \left [
\tr{\sig_k}\Delta^1(\{\sig_k\})\prod_k H(\sig_k\gam t_k),
\tr{\sig_k}\Delta^{-1}(\{\sig_k\})\prod_k H(\sig_k\gam t_k)
\right ]\,,
\eeqn
where the ''min'' in Eq.\ (\ref{eq85}) means that the error probability
corresponds to the smaller fraction of the version space, which belongs to
the minority votes. Note that in (\ref{eq85}) the values $q(\alpha)$ and
$r(\alpha)$ of both order parameters are important.

For the simple perceptron, and for the parity machine in general,
there is a close relationship between (\ref{eq32}) and (\ref{eq85}); we
will return to this point later.

\subsection{Output Noise}

We have investigated the influence of two types of noise. The first one,
called 'output noise', flips the final output with the probability
\beqn\label{eq34}
p_f = \frac{e^{-\bet}}{1+e^{-\beta}}\,.
\eeqn
This corresponds to
\beqn \label{eq33}
\phi(\sig,\{h_k\})=\frac{\exp (-\bet\cdot\delt[-\sig,B(\{{\rm sgn}(h_k)\})] }
{1+\exp(-\bet) }
\eeqn
as probability, to get the output $\sig$ for given fields $\{h_k\}$ at
the hidden units. 
In Eq.\ (\ref{eq33}), we have written $\delt[i,k]$ for the
 Kronecker symbol ($=1$, if $i=k$, $=0$ for $i\ne k$), and
$\beta^{-1}$ is the parameter characterizing the noise strength. 

\subsection{Input noise}

The second
 type of noise, called 'input noise', causes
a noise-perturbation of the local fields at the hidden units, 
$h_k\to h_k+\eta$,
where $\eta$ is a random variable with the Gaussian distribution $\rho(\eta)
=(2\pi\gam)^{-1/2}\exp(-\eta^2/(2\gam))$.
Similarly to the case of output noise,
 it is natural to define the noise strength
$\beta^{-1} :=\gamma$.

The origin of this noise can be that the input pattern itself is
subjected to corruption by noise or that there is weight noise in
the couplings of the teacher.

The flip-probability $p_f(\gam)$
depends here on the architecture, namely it is
\beqn\label{eq40}
p_f(\gam)=\eps\left (\sqrt{(1+\gam)^{-1}}\right )\,\,,
\eeqn
with the generalization error $\eps (r)$ already known from Eq.\ (\ref{eq32}).

The probability $\phi$, to answer on $\{ h_k \}$ with $\sig$, is
\beqn\label{eq45}
\phi(\sig,\{ h_k \} ) =
\tr{\sig_k} \Delta^\sig ( \{ \sig_k \} ) 
\prod_{k=1}^K H(-\sig_k\, h_k\,\sqrt{\beta})\,,
\eeqn
where $H(x):=\int_x^\infty {\rm d}u\exp (-u^2/2)/\sqrt{2\pi}$.

Eq.~(\ref{eq40}) can be seen as follows:
The generalization error $\eps(r)$ defines the probability that student and
teacher machine, which have overlap $r$, give a different answer on a
question. The local fields at a node $k$ of the  respective machines can be
written as $h^t=t$, $h^s=t\cdot r + v \cdot \sqrt{1-r^2}$, with two
independent, normally distributed random variables $t$ and $v$. The
flip-probability, for comparison, defines the probability that the local field
$h^0$, by adding noise with average 0 and variance $\gam$, is changed to
$h^1$ , where $h^0=t$ and $h^1=t+v\sqrt{\gam}$, 
such that the final answers differ.
Thus the problems are completely analogous, with the exact correspondences
\beqn\label{eq41}
\frac{\sqrt{1-r^2}}{r} \,\widehat = \,\sqrt{\gam},\quad{\rm or}\quad r 
\,\widehat = \, \sqrt{\frac 1 {1+ \gam}} \,.
\eeqn

 With (\ref{eq40}), one can easily discuss the small-noise limit in this case,
 since for $r\to 1$, according to Eq.\ (33) in \cite{l:scho1}, one gets
\beqn\label{eq42}
 \eps(r)\to
n_c\pi^{-1}\sqrt{2(1-r)}\,,
\eeqn which follows e.g.\ from Eq.\ (\ref{eq32}) for
$r\to 1$; thus one obtains
\beqn\label{eq43}
p_f(\gam\to 0) \,\,\to\,\, n_c\cdot \pi^{-1}\sqrt{\gam}\,.
\eeqn

Here $n_c$ is, in the limit considered, the only architecture-dependent
value determining the asymptotics $\eps(\gam\to 0)$. It characterizes the
'border-regime' of the Boolean function $B(\{\sig_k\})$, namely by
\beqn\label{eq44}
n_c=\left(\frac{1}{2}\right)
^K\,N_c\,,
\eeqn
where $N_c$ is the number of all those possible $K\cdot 2^K$ bit- flips of
the outputs of the hidden units, which would lead to a change in the final
output, see Eq.\ (35) in \cite{l:scho1}.

\section{Teacher and student machine have identical noise-levels}

In the following we assume at first $\phi_s\equiv\phi_t$, which means that
the student machine uses the known noise-level of the teacher machine.
This assumption, which implies $r=q$,
 is natural, since in this way overfitting will be avoided;
moreover, as we will see later, it is not too far
from the optimal choice.

\subsection{Free energy}

After the preparations in the last section we can calculate the free energy 
for both noise types. One gets from Eqs.\ (\ref{eq27}) and (\ref{eq28})
\beqn\label{eq49}
{\cal F} = {\rm extr}_{(q)}\left \{
 \frac {1}{2} \ln (1-q) + \frac{q}{2} -\al\cdot W(q)\right \}\,,
\eeqn
where for the case of output noise
\beqa \label{eq50}
W(q) = &-&\int \prod_{k} {\rm D}t_k \sum_{\sig} \frac{1}{1+\exp(-\bet )}
\tr{\sig_k} \widetilde\Delta^\sigma(\{ \sig_k \})
\prod_k H(\sig_k\gam t_k) \nonumber \\ &\times& \hspace{1cm} \ln [
\tr{ \sig_k } \widetilde\Delta^\sigma(\{ \sig_k \} )
\prod_k H(\sig_k\gam t_k)]
\eeqa
with
\beqn\label{eq39}
\widetilde\Delta^\sig ( \{ \sig_k  \}) := \Delta^\sig ( \{\sig_k  \})
+ \exp(-\bet )\Delta^{-\sig } ( \{ \sig_k  \} ) 
\eeqn
and for input noise
\beqn \label{eq51}
W(q) = -\int \prod_k {\rm D}t_k \sum_{\sig}
\tr{\sig_k} \Delta^\sigma(\{ \sig_k \})
\prod_k H(\sig_k\bar\gam t_k) \ln %\left 
[
\tr{\sig_k} \Delta^\sigma(\{ \sig_k \})
\prod_k H(\sig_k\bar\gam t_k)%\right
 ]\,
\eeqn
with
\beqn\label{eq48}
\bar \gam =\sqrt{ \frac{q}{1-q+(1/\beta )} }\,.
\eeqn
%Here  $H(t)=\int_{t}^\infty{\rm D}u=\int_{t}^\infty\exp(-u^2/2){\rm
%d}u/\sqrt{2\pi}$.
These formulae will be generalized below for the cases of different
noise-levels of the student and of the teacher machine, and for 1-step
replica symmetry breaking (RSB1).

\subsection{Identical noise-levels: Results for the case of output noise}

\subsubsection{Small training set: $q\to0$}

The limit of a 'small' training set is best characterized by the corresponding
limit $q\to 0$; in some cases  this implies $\al\to 0$, but sometimes the 
corresponding limiting $\al$ can have a finite values as well (see below). 
The opposite limit $q\to1$, which implies $\alpha\to\infty$, will be
considered in the next subsection.

For $q\to0$ a redefinition of the {\it
correlation moments} $a_m^\sig$ of Eq.\ (47) in \cite{l:scho1} suffices to
capture the influence of the noise considered. With
\beqn\label{eq52}
a_m^\sig = \left(\frac{1}{2}\right)^K \tr{\sig_k}\Delta^\sig
(\{\sig_k\})\prod_{i=1}^m\sig_i
\eeqn
one defines
\beqn\label{eq53}
b_m^\sig := \frac
{a_m^\sig +e^{-\beta} a_m^{-\sig} }
{1+e^{-\beta} } % \nonumber \\  &=&
= \cases{a_m^\sig \tanh(\frac{\beta}{2}), &for $m \ge 1 $\cr
 a_0^\sig \tanh(\frac{\beta}{2}) +\frac{e^{-\beta}}{1+e^{-\beta}},&for $
 m=0$\cr}\,\,\,\,\,\,\,\,.
 \eeqn
Therefore, the $b_m^\sig$ fulfill the same algebraic relations, Eq.\ (51) in
\cite{l:scho1}, as the $a_m^\sig$, namely
\beqn\label{eq54}
b_m^\sig = -b_m^{-\sig}, \quad {\rm and}\quad b_0^\sig +b_0^{-\sig}=1 \,.
\eeqn
In particular, the so-called {\it order-index} $n$, see below, is unchanged;
$n$ is defined by
\beqn\label{eq55} b_n^\sig \ne 0,\quad b_m^\sig =0 \,\,\,{\rm for}\,\,\,
1\le m<n\,\,.
\eeqn
Moreover, for $W(q)$ one gets  to lowest order in $q$
\beqa\label{eq56}
W(q) &=& -b_0^1 \ln b_0^1 -b_0^{-1} \ln b_0^{-1} -\ln (1+e^{-\bet})
-\frac{q^n}{2}\left (\frac{2}{\pi}\right )^2 { K\choose n
}\,\,\frac{(b_n^1)^2}{b_0^1b_0^{-1}} +... \nonumber \\
 &=:& w_0 -q^n w_1+...
\,,
\eeqa
which agrees completely with Eq.\ (54) in \cite{l:scho1} apart from the
replacement of $a_m^\sig$ by $b_m^\sig$ and by the non-essential additional
term $\ln(1+e^{-\beta})$. So the results for $q(\al)$ in \cite{l:scho1} can
be simply generalized by these replacements. Only with the generalization error
$\eps (\al)$ we have to keep in mind that after training the noise is switched
off, such that for $\eps(\al)$ the $a_m^\sig$ must be kept.

Concerning the order-index $n$, we thus can state as in \cite{l:scho1}
that
\begin{itemize}
\item for $n=1$, e.g.\ for the committee machine, the overlap $q(\al)$
increases for $\al \ll 1$ proportional to $\al$, i.e.\ there is
generalization right from the beginning, and one gets

\beqn\label{eq57}
q(\al)\to 2\al \frac{K}{\pi}\frac{(b_1^1)^2}{b_0^1b_0^{-1}}\,\,.
\eeqn
This case happens for 6 of the complete set of 9 examples with $K=4$
hidden units given in Fig.\ 1 of \cite{l:scho1}.
\item For $n=2$, phase transitions of second order or of first order (or
both) are possible, as discussed in detail in section 7.2 of \cite{l:scho1}.
Generally, for $n=2$ the network is purely guessing, i.e.~the
error probability is 1/2, as long as
$\al$ is between 0 and the first critical value, whereas an increase of $\al$
beyond this critical value leads to a continous (resp.\ discontinous)
increase of the generalization ability in the case of a 2nd-order (resp.\
1st-order) transition. These transitions with $q(\al)\equiv 0$ for $\al <
\al_c$ are called by us 'Aha-effect transitions'. As just stated, they
appear only for
$n\ge 2$, whereas for $n=1$ only so-called 'interim transitions' (if at all)
can happen: At an 'interim transition' $q(\al)$ is finite already below
$\al_c$. 

If for $n=2$ the 2nd-order transition is not preceded by a 1st-order one,
the critical loading is
\beqn\label{eq58}
\bar\al_c = \frac
{\pi^2(b_0^1b_0^{-1})}{4K(K-1)(b_2^1)^2}\,\,.
\eeqn
The case $n=2$ happens two times in Fig.\ 1 of \cite{l:scho1}.
\item For $n\ge 3$, as in the noise-free case, one always gets a first-order
transition at a critical $\al_c > 0$. Nevertheless this $\alpha_c$ has to
be obtained numerically since the behaviour around a $q=q_c$ with $q_c>0$
is relevant.

This case happens e.g.\ for the
parity machine with $K\ge 3$, which has $n=K$.
% A complete set of nine examples with
% $K=4$  is given in  Fig.\ 1 in \cite{l:scho1}.
\end{itemize}

Thus, as long as there is no transition of first order, the behaviour can be
described analytically by looking how the noise strength $\beta^{-1}$ changes
the correlation moments $b^\sigma_m$. Some of the following statements are
based on this fact; they are not exact as far as the {\it locations}
of first order transitions are concerned, but we do not always state
this limitation explicitly.

If one considers only machines, which have the same probability for the
two possible outputs $\sig=\pm 1$, these results can be simply condensed
into a rescaling 
\beqn\label{eq59}
\al \to\al_{{\rm{\scriptsize eff}}} :=\al\cdot \tanh^2
 \left( \frac{\beta}{2}\right ) \equiv \al\cdot (1-2p_f)^2\,,
\eeqn
where $p_f$ is the flip-probability defined in Eq.\ (\ref{eq34}).
This can be intuitively understood as follows
\begin{itemize}
\item $p\cdot(1-2p_f)$ is the ''uncorrupted fraction'' of the training set.
\item The results are affected by noise from both the teacher and the student
machine, which explains the power of 2 in Eq.\ (\ref{eq59}).
\end{itemize}

\subsubsection{Identical noise-levels; output noise;
large training set: $q\to1$}
In this limit, the teacher
network is approximated with arbitrary accuracy, for every
noise strength $\beta^{-1}$.
 That this is possible, is not at all self-evident:
The training set contains mistakes since the teacher makes errors, the
student learns this set making errors as well, but in spite of these facts 
the teacher machine is
approximated perfectly. As we will see below, a bad training strategy could
empede generalization; so the fact that the student accepts the noise level
of the teacher, as assumed at present, is already a good strategy, although
it is not yet optimal, as we will see later.

From Eq.\ (\ref{eq50}) one obtains for $q\to 1$ the information gain
\beqn\label{eq60}
W(q)\to p_f [\ln (1-p_f) -\ln p_f ]+n_c w_1(\beta)\sqrt{1-q}
\eeqn
with
\beqa\label{eq61}
w_1(\beta)&=&-\sqrt{\frac{2}{\pi}} \int_0^\infty {\rm
d}u \,\, \Bigg\{ \,\, 
\frac{1-e^{-\beta}}{1+e^{-\beta}}\,H(u) \ln \left [ 
\frac{H(u)+e^{-\beta}H(-u)}{H(-u)+e^{-\beta}H(u)}   \right ]
\nonumber \\
&&+\frac{e^{-\beta}}{1+e^{-\beta}}\ln \,[e^\beta H(u)+H(-u) ] +
\frac{1}{1+e^{-\beta}}\ln \,[e^{-\beta} H(u)+H(-u) ]
 \,\,\Bigg\}\,.
\eeqa
Determining $q(\al )$ from $W(q)$ and inserting the result again into Eq.\
(\ref{eq42}), one obtains
\beqn\label{eq62}
\eps (\al )\to\frac{\sqrt{2}}{w_1(\beta)\pi\,\al}\,.
\eeqn
So again, as in the deterministic case, one has an $1/\al$-asymptotics, and
the prefactor does not depend on the Boolean function $B(\{ \sig_k \})$, but
only on the noise-level $\beta$. Therefore again, one is lead to a rescaling
\beqn\label{eq63}
\al \to\al_{{\rm{\scriptsize eff}}} :=r(\beta )\cdot\al \equiv\frac{w_1(\beta)}
{w_1(\infty)}\cdot\al\equiv\frac{w_1(\beta)}{0.720647}\cdot\al\,.
\eeqn

In Fig.~2, the scaling parameter $r(\beta )\equiv w_1(\beta)/0.720647$,
which applies to the regime $q\to 1$, and the ''intuitive'' scaling
parameter $r^{it}(\beta):=[1-2p_f(\beta )]^2$, which applies to the limit
$q\to 0$, are presented as a function of the ''flip parameter'' 2$p_f$,
which corresponds to the ''corrupted fraction of the training set''. Obviously
$r(\beta)$ is $< r^{it}(\beta)$, which means that for large loading the
effect of noise is stronger than for low loading. But the difference is not
large, so that the ''intuitive reduction factor'' $r^{it}(\beta )$
 always will give a good estimate of the effect. But it should be noted that
in the limit $q\to 1$ already a slight corruption of the training set leads
to a significant reduction of the generalization ability because of the
infinite slope of
$r(2p_f)$ in the limit $p_f\to 0$.
%\{Data collapsing}

\subsubsection{Output noise: Data collapsing in the large-noise limit}

Strong noise is described by $\beta\to0$. Calculating the behaviour in
this limit for both, $q\to 0$ and $q\to 1$, one sees that $q(\al,\beta )$ is
$\propto \beta^2\al$.
We have looked for (approximate) data-collapsing in the whole
parameter region  $0.2 < \beta < 1$, by plotting the curves $q(\al, \beta)$
not only as a function of $\al$, with $\beta$ as curve parameter, but using,
instead, also the product $\beta^2\al$ as scaling variable. The results are
compared in Fig.\ 3 and show that with the variable
$\beta^2\al$, for $\beta \le 1$, a good, although still approximate,
data-collapsing is obtained in the whole region $0 < \al\beta^2
 <\infty$. This data-collapsing becomes asymptotically exact in the
 above-mentioned  limits $q\to 0$ and $q\to 1$.

\subsection{Identical noise-levels: Results for the case of input noise}
\subsubsection{Small training set: $q\to 0$}
Again we consider at first the limit $q\to 0$. In this case one gets
from Eq.\ (\ref{eq48})
\beqn\label{eq64}
\bar\gamma \to \sqrt{q\cdot\frac{1}{1+\gam}} =: \sqrt{q\cdot\zeta}\,,
\eeqn
and for the free energy
\beqn\label{eq65}
-{\cal F}\to {\rm extr}_{(q)}\left \{
 -\frac{q^2}{4}-\al\cdot (w_0-q^n\zeta^n w_1)\right \}\,,
\eeqn
where $w_0$ and $w_1$ are defined as
\beqn\label{eq66}
w_0=-a_0^1\ln a_0^1 -a_0^{-1}\ln a_0^{-1},\quad %{\rm and}\quad
w_1=\frac{1}{2}\,\left (\frac{2}{\pi} \right )^n {K\choose
n}\,\,\frac{(a_n^1)^2}{a_0^1a_0^{-1}}\,.
\eeqn
Again, the order-index $n$, and thus the {\it qualitative} behaviour,
remains unchanged by the noise. Concerning the three cases of $n$, we have
now
\begin{itemize}
\item For $n=1$, the $q(\alpha\to0)$ is given by
\beqn\label{eq67}
q(\al)\to 2\al\zeta\frac{K}{\pi}\frac{(a_1^1)^2}{a_0^1a_0^{-1}}\,\,.
\eeqn
Since $\zeta <1$, the noise diminishes the overlap.
\item For $n=2$, the critical value $\bar\al_c$ of the 2nd-order phase
transition (if it is not preceded by a 1st-order one, see above) shifts to
a higher value:
\beqn\label{eq68}
\bar\al _c(\zeta) =\frac{1}{\zeta^2}\bar\al_c(0)\,.
\eeqn
\item For $n\ge 3$, there is again a 1st-order phase transition, but the
resulting critical loading must be determined numerically, since the
behaviour around a $q=q_c$ with $q_c>0$ is relevant.
\end{itemize}
Taking all three cases together, we find a rescaling, which also applies for
$n\ge 3$, namely
\beqn\label{eq69}
\al\to\al_{{\rm{\scriptsize eff}}}=\zeta^n\cdot\al\,.
\eeqn 
We remind again on the caution which has to be taken as far as first order
transitions are concerned.

\subsubsection{Identical noise-levels; input noise; large training sets:
 $q\to1$}

Considering the limit $q\to 1$, i.e.\ $\al\to\infty$, one sees
from Eq.\ (\ref{eq48}) that now $\bar \gamma$ does not converge to
$\infty$, in contrast to the behaviour {\it without} noise, or with {\it
output} noise.
Instead
\beqn\label{eq70}
\bar\gamma\to \sqrt{\beta}\,[1-\frac{1}{2}(1+\beta )(1-q)]\,.
\eeqn
 Expanding Eq.\ (\ref{eq50}) for $1-q\to 0$, one obtains
\beqn\label{eq71}
W(q)\to w_0(\bet) + (1-q)\cdot w_1(\bet)\,.
\eeqn
If one determines $q(\al)$ from Eq.\ (\ref{eq31})
 and inserts it into Eq.\ (\ref{eq42}), one
obtains
\beqn\label{eq72}
\eps(\al)\to n_c\,\frac{1}{\pi\sqrt{w_1(\beta)\,\al}}\,,
\eeqn
with
\beqa\label{eq73}
w_1(\beta) &=& \frac{(1+\beta)\sqrt{\beta}} {2\sqrt{2\pi}}
\int \prod_k {\rm D}t_k\sum_{\sig=\pm 1} \tr{\sig_k}
\Delta^\sig(\{\sig_k\})%\nonumber \\
\left [\sum_m\sig_mt_me^{-\beta t_m^2/2}
\prod_{k(\ne m)}H(\sig_kt_k\sqrt{\beta})\right ]\nonumber \\
&\times&\ln\left [\tr{\sig_k}\prod_{k(\ne
m)}H(\sig_kt_k\sqrt{\beta})\right ]\,.
\eeqa
From Eq.\ (\ref{eq72}) one can see that  {\it input noise}, in contrast to
output noise, leads to a drastic deterioration of the generalization
ability, namely from an asymptotics as $\eps(\al)\to c/\al$ to the slower
decrease $\eps(\al)\to \tilde c /\sqrt{\al}$. Additionally we find that  the
prefactor $\tilde c$ of this behaviour -- in contrast to $c$ -- depends on the
architecture. In Fig.\ 4, for the $K=2$- and $K=3$-parity machines and for the 
$K=3$-committee, we present this {\it prefactor} $c^{~}$ of the asymptotic
behaviour $\eps(\al\to\infty)\to  \tilde c /\sqrt{\al}$ as a function (i) of
$\gamma$ and (ii) of the ''flip probability''
$p_f=n_c\pi^{-1}\sqrt{\gamma}$.  Interestingly, with the last-mentioned
representation, the data almost collapse to a single curve, although the
results look quite different when presented against $\gamma$.

For the simple perceptron a corresponding result is given in 
\cite{l:opperkinzel}.

\subsubsection{Input noise: Data collapsing in the large-noise limit}

For the case of input noise it can further be shown that 
in the high-temperature limit $\beta \to 0$ and for $\al \to \infty$
\beqn\label{eq76}
\eps(\al)\to n_c\frac{\sqrt{2}}{\pi}\frac{1}{\sqrt{V_0\beta^n\al}}\,,
\eeqn
with
\beqn\label{eq77}
V_0=\left ( \frac{2}{\pi}\right )^n \sum_{\sig =\pm
1}\frac{(a_n^\sig)^n}{a_0^\sig} {K\choose n} .
\eeqn
The corresponding behaviour for $\alpha\to0$ can be seen from
(\ref{eq69}). Summarized, data collapsing  in the limit
$\beta\to 0$ both for $q\to 0$ and $q\to 1$, but with $n$-dependent rescaling
 $\al\to\al_{{\rm{\scriptsize eff}}} =\beta^n\al$.

  In Fig.\ 5, for two examples, we check whether for the input noise
  temperatures $\gamma=1/\beta=1.0$, $2.0$,...,$5.0$ one gets data 
  collapsing. Thereby we use $\zeta=1/(1+\gamma)$ rather than $\beta$ for 
  rescaling which is a better choice if $\beta$ is not close to $0$.
  We compare  results for
  $q$ plotted as a function of $\al$ with results, where $q$ is plotted
  against $\al/\zeta^2$ for the $K=2$-parity machine, resp.\ against
  $\al/\zeta$ for the K4\_mcode2 machine.

\subsection{Identical noise-levels:
The different impact of output- and input noise}

As already seen there are some significant differences of the impact caused 
by output- and input noise, respectively. Let us first have a closer look 
on the influence of a small disturbance by noise.

\subsubsection{Small-noise limit: $\beta\to\infty$}

To compare the impacts of noise in this limit we have to distinguish between
the cases of small and large reduced size $\alpha=p/N$ of the training
set, respectively.
For small $\alpha$
the effect of input noise is for $\gamma\to 0$ (or $\beta\to\infty$)
according to Eq.\ (\ref{eq69})
\beqn\label{eq74}
\al_{{\rm{\scriptsize eff}}}=\zeta^n\cdot\al=(1+\gamma)^{-n/2}\cdot\al
\to (1-\frac{n}{2}\gamma)\,\al \,.
\eeqn
With the flip rate $p_f$ given by Eqn. (\ref{eq43}) in this limit we have
for {\it input noise}
\beqn\label{eq75}
\al_{{\rm{\scriptsize eff}}} = \al\cdot  [ 
1-\frac{n}{2}\left (\frac{\pi p_f}{n_c}\right )^2    ]\,.
\eeqn
The training set is therefore only reduced by a small amount $\propto p_f^2$.
In contrast, for {\it output noise} this amount is $\propto p_f^1$. For
machines with equal probability for final output $\sigma=\pm1$ this can
directly be seen from Eqn. (\ref{eq59}). 
This means that
a small amount of input noise does hardly matter for the case of
$\al\to 0$, in contrast to a small amount of output noise, which -- so to
say -- instantly deteriorates the behaviour.

On the other hand, for $\al\to\infty$ a small amount of input noise induces
a {\it qualitative change} in the asymptotics, since then 
the behaviour is shifted from the
$1/\al$-asymptotics to the slower $1/\sqrt{\al}$-decrease, 
if $\al $ is beyond the corresponding (non-universal) crossover value.
 In contrast, for output noise the $1/\alpha$-behaviour is qualitatively
 unchanged,
only the prefactor is increased.

\subsubsection{Input noise: Disappearance of phase transitions}

A further difference concerns the phase transition
of the K4\_mcode2 machine: Here the {\it intermediate}  phase transition
as existent in the deterministic case (see \cite{l:scho1}) disappears
for the case of input noise. This does not happen  for output noise.
Whether other intermediate phase transitions are affected similarly has
to be checked.

Nevertheless all other types of phase transitions (see subsection 3.2.1
above or \cite{l:scho1}) persist for {\it both cases} of noise, although
the  shift of the critical parameters depends of course both on the
strength and on the type of the noise.

\section{Optimization of the noise-level of the student machine}

In this section we focus exclusively on {\it output noise}, but now
different noise-levels  for student and teacher networks are allowed.
The following two sections \ref{secrs} and \ref{secrsb} give the theory
for the general multilayer case. The formulas are evaluated mainly for
the simple perceptron; for the general case just asymptotic results
within the replica symmetric formalism are given.

\subsection{Different
 noise-levels; output noise: Replica-symmetric formalism}
\label{secrs}
In the replica-symmetric formalism of the preceding section one has now two
different noise strengths $\beta_t$ and $\beta_s$ of the teacher resp.\ the
student machine, and additionally now $q\ne r$. Therefore, instead of Eqs.\
(\ref{eq49}) and (\ref{eq50}) one has
\beqn\label{eq78}
{\cal F}={\rm extr}_{(q,r)}\left \{
 \frac {1}{2} \ln (1-q) +\frac{q-r^2}{2(1-q)} -\al\cdot
W(q,r)\right \} \,,
\eeqn
where for the present case of output noise it is
\beqa\label{eq79} 
W(q,r) &=& -\int \prod_k {\rm D}t_k \sum_{\sig} \frac{1}{1+\exp(-\beta_t )}
\tr{\sig_k} \widetilde\Delta^\sigma(\{ \sig_k \})
\prod_k H(\sig_k\gam_r t_k) \nonumber \\ && \hspace{1cm}\ln [
\tr{\sig_k} \widetilde\Delta^\sigma(\{ \sig_k \} )
\prod_k H(\sig_k\gam t_k)]
\eeqa
with 
\beqn\label{eq80}
\gamma =\sqrt{\frac{q}{1-q}}, \quad{\rm and}\quad \gamma_r
=\frac{r}{\sqrt{q-r^2}}\,,
\eeqn
and with $\widetilde\Delta$ defined by Eq.\
(\ref{eq39}) with $\beta\equiv\beta_t$ and $\beta_s$, respectively.
The functions $q(\al)$ and $r(\al)$ follow again from the saddle-point
conditions (\ref{eq31}). An additional quantity of interest is the {\it
relative training-error} $\eps_{tr}:=E_{tr}/p=-\al^{-1}\partial{\cal F}/\partial
\beta_s$; $\eps_{tr}$ is thus the fraction of the training set, which is
misclassified by the student machine. A straightforward calculation, see
\cite{l:schottkydiss}, yields
\beqn\label{eq81}
\epsilon_{tr}=\frac{1}{1+e^{-\beta_t}}\int\prod_{k=1}^K{\rm D}t_k
\sum_{\sigma=\pm1}
\frac{\tr{\sigma_k}\widetilde\Delta^\sigma(\{\sigma_k\})
\prod_kH(\sigma_k\gamma_rt_k)}
{\tr{\sigma_k}\widetilde\Delta^\sigma(\{\sigma_k\})\prod_k
H(\sigma_k\gamma t_k)}
e^{-\beta_s}\tr{\sigma_k}\Delta^{-\sigma}(\{\sigma_k\})\prod_k
H(\sigma_k\gamma t_k)\gp
\eeqn
\subsection{Different  noise-levels; output noise: Replica symmetry
breaking} \label{secrsb}
 Within the usual 1-step replica symmetry breaking (RSB1) scheme, see
\cite{l:mezard}, one gets with $q^{a,b}\equiv 1$ for $a=b$,\,\,
$q^{a,b}\equiv q_1$, if $a$ and $b$ are different, but belong to the
same subgroup of $m$ of the $n$ replicas, and $q^{a,b}\equiv q_0$ else,
and with $\Delta q :=q_1-q_0$\ :
 \beqn\label{eq82} {\cal F}={\rm
extr}_{(r,q_0,q_1,m)}\left \{ \frac{q_0-r^2}{2(1-q_1+m\Delta q)}+ \frac
{1}{2} \ln (1-q_1)+\frac{1}{2 m}\ln \left (1+\frac{m\Delta
q}{1-q_1}\right ) -\frac{\al}{m}\cdot W(r,q_0,q_1,m) \right \}\,, \eeqn
with
\beqa\label{eq83} W(r,q_0,q_1,m)&=& -\int \prod_k {\rm D}t_k
\sum_{\sig} \frac{1}{1+\exp(-\beta_t )} \tr{\sig_k }
\widetilde\Delta^\sigma(\{ \sig_k \}) \prod_k
H(\sig_k\frac{r}{\sqrt{q_0-r^2}} t_k) \nonumber \\ &\times& \ln \left \{
\int \prod_k{\rm D}v_k \left ( \tr{\sig_k } \widetilde\Delta^\sigma(\{
\sig_k \} ) \prod_k H\left [\sig_k\frac{t_k\sqrt{q_0}+v_k\sqrt{\Delta
q}} {\sqrt{1-q_1}} \right ]\right )^m \right \}\,. \eeqa
\subsection{RS and RSB results for the simple perceptron}
 In the following we use the notation 'perfect student' or 'perfect
learning' to describe the fact that the given training set $\cal TS$ is
learned {\it without} any error, so '${\cal{TS}}$-perfect' would be a more
precise terminology. 'Non-perfect learning' (or better '$\cal TS$-nonperfect
learning') means that errors with respect to $\cal TS$ are made. We stress
at this place that $\cal TS$ may already be corrupted with respect to
the original {\it rule}. (We also mention that 'perfect learning' is
sometimes as well used to describe coinciding architecture of student and
teacher which is here the case anyway. To keep our special definition in
mind,  we always use primes in the terminology
'perfect'.)

\subsubsection{RS results:
'Non-perfect' teacher, but 'perfect' student}
In the following, for simple perceptrons ($K=1$) we consider at first the
case of a 'perfect' student machine (i.e.\ necessarily $\beta_s=\infty$)
 but allow for a
'non-perfect' teacher ($\beta_t < \infty$), which means that the
training set itself is partially corrupted, since the answers given by
the teacher on the questions $\xib^\mu$ ($\mu=1,...,p=\alpha\,N$) do
not always follow the rule, but are partially random.

Of course, 'perfect learning' of the {\it corrupted} training set is then
possible only up to a specific $\alpha_c$ depending on the noise; e.g.~if
for {\it all} input patterns $\xib^\mu$ the outputs, prescribed by $\cal
TS$, would be randomized with respect to the original rule, then one would
get the famous result $\alpha_c=2$ of E.~Gardner, \cite{l:gardner}.

 In Fig.\ 6a, both for the case of output noise  and for input noise,
$\al_c$ is presented as a function of the non-corrupted fraction
$(1-2p_f)$ of the training set, with $p_f$ taken from Eqs.~(\ref{eq34})
and (\ref{eq40}), respectively. Additionally, in Fig.\ 6b, for output
noise with $T_t=1$, the curve $r(\al)$ for {\it Maximal Stability
Learning} (MSL) is shown (the solid line) and compared with the
corresponding result for Gibbs learning (the dashed line). For the MSL
case, the student is not a random member of the version space but has
those couplings, which lead to {\it maximal stability} of the
classification of the whole training set of patterns mapped to $+1$ and
$-1$, respectively. This specific member can be obtained by the
well-known AdaTron algorithm, \cite{l:Anlauf}.

Two overfitting effects can be seen: Considering Gibbs learning, the
overlap {\it decreases} at the end of the curve. Compared to MSL, Gibbs
learning is worse for small $\alpha$ (which is expected) but becomes
better for  $\alpha\to\alpha_c$, although MSL chooses a specific vector
as student which is supposed to perform the classification task very
well.

These are hints that training with noise might avoid
overfitting effects of the student and therefore could lead to an
enhanced performance\ :

\subsubsection{RS results:
'Non-perfect' teacher and 'non-perfect' student}
Here we consider again the perceptron ($K=1$) and assume a given output
noise-level $\beta_t=1$ of the teacher machine.

In Fig.\ 7 we present the dependence
of various quantities on the noise strength $T_s :=1/\beta_s$ of the student
machine. These are
\begin{itemize} 
\item The overlap $r(T_s)$ of the two coupling sets. This quantity
determines the generalization ability of the Gibbs training algorithm.
\item The typical overlap $q(T_s)$ of two student perceptrons.
\item The overlap $r_{{\rm\scriptsize{cp}}}$ of the so-called
''central-point network'' with the teacher machine: The coupling vector of
the ''central-point network'' is obtained as a (weighted) average of the
coupling vectors of all student perceptrons forming the Gibbs ensemble.
Analogously to Eqs.\ (75) and (76) in \cite{l:scho1} one can show that
$r_{{\rm\scriptsize{cp}}}(T_s)=r(T_s)/\sqrt{q(T_s)}$.
The corresponding generalization error is obtained by plugging 
$r_{{\rm\scriptsize{cp}}}$ into (\ref{eq32}).

Moreover, for $\beta_s=\beta_t$ the corresponding network turns out to have
the same generalization ability as an exploition of the version space
performed by the Bayes algorithm, see \cite{l:scho1}.
 
\end{itemize}
From the results of Fig.\ 7, the following points should be noted\ :

1. For all values of $\al$, $r(T_s) $ and $q(T_s)$ cross at $T_s=T_t$. In
fact, for this case the teacher has the same properties as a typical student
of the Gibbs ensemble.

2. The curves $r_{{\rm\scriptsize{cp}}}(T_s)$ have a flat maximum at
$T_s=T_t$. This is also obvious: Since in this case the central-point network
reaches the generalization ability of the Bayes algorithm, which is maximal, 
according to information theory.
Nevertheless, since the curve $r_{{\rm\scriptsize{cp}}}(T_s)$ is very 
flat around the maximum, the detailed value, $T_s=T_t$, is non-essential.

3. In contrast, the overlap $r(T_s)$ for Gibbsian learning shows a maximum
for a finite noise level $T_s$ only for $\al=2$ and $\al=5$, but not for
$\al=1$. Obviously for Gibbs learning, training with output noise (i.e.\
$T_s > 0$) is only advantageous beyond a finite $\al$, which of course
depends on $T_t$. For a similar model this was reported already in
\cite{l:gyoergyi}.

Fig.\ 8 presents the optimal value of the student noise-level $T_s$ as a
function of $\al$ for fixed $T_t=1$. Beyond $\al=\al_c$ ($\approx 0.6$
in Fig.~8) training
with noise leads to better generalization. We determined the value of
$\al_c$ only numerically from the appearance of a maximum. For large $\al$,
the optimal $T_s$ for $T_t=1$ converges to $0.60524$.

In Fig.~9, the training error $\eps_{tr}$ is presented as a function of
$T_s$ for $T_t=1$ and for different $\al$-values, ranging from $\al=0$
(lowest curve for small $T_s$) to $\al\to\infty$. For $T_s\to\infty$ all
curves converge to $\eps_{tr}=0.5$, as they should. For $\al\to 0$ the
result is only determined by $ \beta_s$, and for $\al\to\infty$ only by
$\beta_t$, namely 
\beqn\label{eq84}
\eps_{tr}(T_s)_{|\al\to 0} =\frac{e^{-\beta_s}}{1+e^{-\beta_s}}\,,\quad
\eps_{tr}(T_s)_{|\al\to \infty} =\frac{e^{-\beta_t}}{1+e^{-\beta_t}}\,.
\eeqn

\subsubsection{Different noise-levels; simple perceptron: RSB results}
The non-monotoneous behaviour of $r(T_s)$ for $\al=2$ in Fig.\ 7
is not an artefact of replica symmetry: For $\al=2$  we
determined an optimal $T_s > 0$, and since the problem is {\it learnable}
for this $\al$ and every $T_t$, replica symmetry is correct.

However, when $\al$ becomes $ > \al_c$ (which is always larger
than 2 according to Fig.\ 6a), one expects replica
symmetry breaking (RSB).
In Fig.~10, for $\al =5$ and $T_t=1$, the order parameters $r(\al)$ and
$q(\al)$ are
presented, as obtained in RS (the bold line) and RSB1 theories. Obviously,
replica symmetry is correct around the optimal student temperature
$T_s\approx 0.35$ and beyond, but for smaller values ($T_s\lsim 0.27$) replica
symmetry is broken. In fact, in the RS approach, a paper of Gy\"orgyi,
\cite{l:gyoergyi},  predicts (for a similar model) a rather large and 
useful effect of
''training with noise'' already for the case of small $T_s$.
 However, due to RSB corrections, see Fig.\ 11,
the benefit of noise is weaker than expected in the RS calculation of
\cite{l:gyoergyi} since
for the case of small noise of the student machine the overlap
$r^{{\rm\scriptsize{RSB}}}$
calculated in RSB1 is higher than calculated with RS,
and the value obtained in
RSB1 for $\al\lsim 0.27$ is almost as high as that obtained at the
optimal value $T_s\approx 0.35$, where RS is correct.  An extension of
replica-symmetry breaking e.g.\ to RSB2 (see \cite{l:mezard}) may even 
enhance the value of $r(T_s)$ calculated in RSB1.

 At this place we mention that T.~Uezu, in a recent preprint,
\cite{l:Uezu}, has independently treated  problems studied in this
section 4.3, with largely overlapping results.
 
\subsubsection{Why RSB occurs: An intuitive explanation}

As stated above, RSB can occur if $\al$ becomes $ > \al_c$ (which is always 
larger than 2 according to Fig.\ 6a). Nevertheless RS is restored if
the student temperature $T_s$ increases above a critical point as 
demonstrated in the previous paragraph.

To understand this, let us consider the case $T_s\to 0$, which means that
only student machines with minimal training error are allowed. The available
phase space will then separate into disjunct parts: To every disjunct part
student machines belong, which misclassify a certain (different) minimal set
of patterns of the training examples. The student machines ''in-between the
disjunct components'' make more errors: If $T_s$ is increased, they also
become ''more and more allowed'' until finally the allowed region of phase
space melts together to a single component, i.e.~replica symmetry is
restored. This scenario is presented qualitatively in Fig.\ 12.

\subsection{Different noise-levels; output noise: 
RS and RSB results for multilayer networks}

Since the numerical effort increases drastically for multilayer network
we consider exclusively output noise and restrict ourselves to
give just asymptotic results for this case.  For $T_t=0.2$, $0.5$, $1.0$
and $2.0$ the ratio $T_s/T_t$ is varied.

\subsubsection{Large training sets: $q\to 1$}
We ask, how the student temperature $T_s$ should be chosen in the case
of large training sets. The aim is to obtain an optimal prefactor
for the asymptotic behaviour of the generalization error, 
$\eps(\al)\to c_0/\al$. This can be calculated analytically for given $T_t$,
results are presented in Fig.\ 13.
There, for 4 different values of $T_t$ we present results for the ratio
$c_0/c_0^{{\rm\scriptsize{opt}}}$ of the coefficients $c_0$ of the
asymptotic behaviour $\eps(\al)\to c_0/\al$. Here
$c_0^{{\rm\scriptsize{opt}}}$ refers to the optimal choice of $T_s$ for
given $T_t$. The results have been plotted as a function of $T_s/T_t$.
 Again we find that the optimal choice is $T_s/T_t\approx 0.6$, but with a
''flat behaviour''. Note that these results do not depend on the
architecture, in contrast to corresponding results for input noise.
\subsubsection{Small training sets: $q\to 0$}
The crucial question in this limit is, in which way the above-mentioned
phase-transitions shift when a noise strength $T_s\neq T_t$ is used, in
particular whether there is an optimal $T_s^{\mbox{\scriptsize opt}}\neq T_t$
for which the ''Aha-effect'' happens earlier, i.e.\ for smaller $\al$.

For the {\it parity machine} this question can be answered at once: There
the choice $T_s=T_t$ means that the already mentioned ''central-point
network'' with overlap $r/\sqrt{q}$ reaches the Bayes generalization
ability, which is optimal, i.e.\
$\eps_{{\rm\scriptsize{Bayes}}}(q)=
\eps_{{\rm\scriptsize{Gibbs}}}(\sqrt{q})$.
Therefore, for the parity machine the ''Aha-effect'' phase transition cannot
happen earlier than calculated for $T_s=T_t$.

For other networks, a slight addition to the argument is in order, since now
the generalization ability of the ''central-point network'' is smaller than
that one obtained with the Bayes prescription. But for $T_s=T_t$
  the typical student is also a typical teacher, i.e.\ we have an ensemble
of student coupling vectors, which corresponds to the {\it
a-posteriori-probability} that a certain coupling vector is that of the
teacher machine. The Bayesian generalization error can be calculated from
the expectation value of $q$ for this ensemble by means of (\ref{eq85}).
This implies that the Bayes generalization ability becomes trivial (i.e.~the
error probability is 1/2) below the critical
$\alpha$ calculated for $T_s=T_t$ 
(Since the Bayes generalization is
optimal, a variation of $T_s$ cannot lead to further improvement). So a
transition to nonzero $q$ cannot occur earlier for whatever $T_s$ one
chooses, and thus for all networks considered the phase transition cannot
appear for smaller $\alpha$.

\section{Why is training with noise useful\ ?}
That training with noise can be useful, is already 'folklore', see e.g.\
\cite{l:poeppel} or \cite{l:stroud}, and in the present context it is also
known that in this way one can avoid overfitting (see \cite{l:nehl,l:watkin}).
Here we want to go somewhat more into the details and look into the phase
space structure. 

\subsection{Survey of the phase space of a small system}

Let us first define the so called 'genuine error' $E_0$ of the student:
 This quantity
counts the number of patterns from the training set where the
(deterministic) answer of the student disagrees {\it with the original
rule} and not with the partially corrupted answer of the teacher
network.

Now we perform a simple survey of the phase space of a small perceptron
with $K=1$ and $N=10$, with normalized randomly chosen coupling vector
$\wb^t$ with $N$ components, and with a given set of $p=20$ questions
$\xib ^\mu$, i.e.~we have $\alpha =p/N=2$. By flipping 5 of the 20
answers given by $\wb^t$, we have our teacher perceptron endowed with
an output noise level of $p_f=5/20=0.25$ and with a training set
consisting of the 20 pairs of questions and the partially corrupted
answers. Then a large number of {\it student vectors} $\wb^s$
 are drawn randomly from a uniform prior over all normalized real vectors
with $N=10$, where the random components are sampled from a Gaussian
distribution with zero average and variance $1/N$. Finally, for each
$\wb^s$, we evaluate (i) the error $E_{tr}$ with respect to the actual
training set, i.e.~the corrupted one, (ii) the error $E_0$ with respect to
the {\it uncorrupted} answers on the training patterns, and (iii) the actual
overlap
$r$ with $\wb^t$. The results are condensed into a two-parameter table: For
each combination of the values of $E_{tr}$ and $E_0$ the {\it number} of
vectors (yielding these errors) as well as the corresponding averaged
overlap $r$ is stored.

This table can be used to calculate the $r(T_s)$ curve for the specific
values of $\alpha=2$ and $p_f=0.25$. This curve is shown in Fig.~14 (the
solid line) and compared to the theoretical result with $\al=2$ and
$T_t=0.91$ (corresponding to $p_f=0.25$) which has been evaluated
analogously to Fig.\ 7. We see that there is actually -- in spite of the
smallness of the simulated system -- a nice qualitative similarity (more
would be unexpected) in the behaviour of $r(\al)$ as a function of $T_s$; in
particular we find that it is useful to increase the noise until
$T_s\approx 0.4$, whereas for larger $T_s$ noise is more and more
detrimental.

Looking more directly at the performance of this small system,  we note the
following numbers from a typical realization of the survey through its phase
space: In this survey the combination ($E_0=5$, $E_{tr}=0$) was realized 215
times: These 215 students made no error with respect to the actual
(i.e.~corrupted) training set, and thus they made $E_0=5$ errors with
respect to the original rule. Possible values for $E_{tr}=1$ are then (i)
$E_0=4$, i.e.\ one of the 5 errors of the student with respect to the
original rule has been corrected, or (ii) $E_0=6$, i.e.\ one additional
error has been made. In the simulation we found that (i) happened in 1746
cases, whereas (ii) was less frequent, namely 1146 times, although there are
many more combinatorial possibilities for case (ii). The corresponding
averaged overlaps are of course increased for the case $E_0=4$ and decreased
for $E_0=6$. So obviously the system is able to correct errors previously
made with respect to the original rule, and to increase the typical overlap
$r$ in this way.

\subsection{The error $\epsilon_0$
 with respect to the uncorrupted training set} This can as well be confirmed
theoretically. Defining $\epsilon_0=E_0/p$ as the fraction of errors of the
student with respect to the original training set, one
 derives the following equation~:
\beqn
\epsilon_0=\frac{2}{1+e^{-\beta_t}}
\int {\rm D}t
\frac{H(\gamma_rt)H(-\gamma t)e^{-\beta_s}+H(-\gamma_rt)H(\gamma t)e^{-\beta_t}}
{H(\gamma t)+e^{-\beta_s}H(-\gamma t)}\,\,,
\eeqn
with $\gamma$ and $\gamma_r$ defined in (\ref{eq80}).
In Fig.\ 15, $\epsilon_0$ is plotted as a function of
 $T_s$ for $\alpha=2$ and
$T_t=1$. So with increasing $T_s$, $\epsilon_0$ decreases at first,
which means that at first the system mainly corrects the mistakes
contained in the corrupted training set  with respect to the original
rule. But after a minimum, $\epsilon_0$ increases again for
larger $T_s$ and approaches the 'training error' $\epsilon_{tr}$ (i.e.~with
respect to the corrupted training set) for
$T_s\to\infty$.

The reason for this ability to produce the 'right' errors can be
explained in terms of a sort of energy-entropy competition:

Allowing a given student to make errors there are more possibilities to
increase the number of errors than to reduce this number; nevertheless,
there are many more students in the version space classifying the
training set with a decreased error number than with an increased one.
 
\subsection{Multifractal phase space analysis}
A more thorough analysis of the phase space structure is possible with the
recent 'multifractal technique' of Monasson and Zecchina
(\cite{l:monasson,l:zecchina}). Extending this analysis to problems with
noise, the distribution of the phase space volumina $V_\tau$ weighted
with the corresponding degree of membership corresponding to a set
$\tau=\{\sig^\mu\}_{|\mu =1,...,p}$ of answers on the training questions
$\xib^\mu$ is given by the quantity
 \beqn\label{eq86} g(m)=\frac{1}{mN}\left\langle \ln \sum_\tau\left (
e^{-\beta_sE_{tr}(\tau)}V_\tau \right )^m\right\rangle\,\,. \eeqn
Here $m$ is the inverse of a formal temperature and controls, how strong
$V_\tau$ is weighted. Formally $(-g(m))$ is a free energy derived from
the partition function \beqn\label{eq87} Z=\sum_\tau e^{-mE_\tau} \eeqn
with the energy $E_\tau :=-\ln [e^{-\beta_s E_{tr}(\tau)}V_\tau]$. Then
the quantities
\beqn\label{eq88} -k(m) :=-\frac{\partial (m\,
g(m))}{\partial m}, \quad {\rm and}\quad c(k) :=\frac{\partial
g(m)}{\partial (1/m)} \eeqn
 correspond formally to internal energies and entropies w.r. to $T_m
:=1/m$, and measure, how the phase-space volume $V_\tau$ and the
corresponding number of realizations scale in the thermodynamic limit,
namely as $e^{Nk}$ and $e^{Nc(k)}$, respectively. Here $k< 0$, but $c(k)
> 0$. These quantities can be calculated by a formalism, which resembles
a RSB1 calculation, see \cite{l:monasson,l:zecchina,l:schottkydiss}. The
dominating behaviour, which is already calculated in the usual RS
calculation, $m=1$ , is noted by the asterisks in Fig.\ 16 and follows
from the identity ${\rm d}[c(k)+k]/{\rm d}k =0$. In this Fig.\ 16, the
results are for the simple perceptron: The overlap $r(k)$ is presented
as a function of $-k$, for $\al =5$ and the teacher output noise
temperature $T_t=1$, for different values of $T_s$, namely $T_s=1.0$,
$0.8$, $0.6$, $0.4$, $0.3$, $0.2$, $0.16$, $0.14$, $0.12$, and $0.1$,
from the left. In this way, a differentiated picture of changes in the
phase-space distribution induced by changes of the noise temperature
$T_s$ can be given. So there are two effects: With increasing
temperature, regions with high student-teacher overlap become more and
more active. At the same time, the volume determining the typical
(dominant) overlap, i.e.\ the position of the asterisk, moves towards
the maximum of these overlap curves. After having reached the optimal
temperature, the curve decreases again, and regions with high training
error and small overlap begin to dominate the phase space.

\subsection{Why noise is useful: an intuitive picture}

Summarizing the calculation and results of this section 
one can give an intuitive picture for the observed behaviour:
In Fig.\ 16 we plot a scenario corresponding to a slice through a lake
with a flat bank on the left, but a steep shore on the
right. 

To get the basic point let us think of the simple perceptron in a regime
where replica-symmetry is preserved even for $T_s=0$, thus the (corrupted)
training set is learnable. The deepest point in the lake corresponds to the
 error-free solution(s) with respect to this (corrupted) training set.

An increase of the student temperature $T_s$ corresponds to an increase of
the water level starting from this deepest point.
Of course in the case of Fig.~16 the direction of the flat bank is
favoured compared to the steep shore, so the center of mass shifts to
the left for increasing water level.

This can be transfered to our model:
\begin{itemize} 
\item The direction of the flat bank (on the left) corresponds 
to noise-induced flips in the answers which {\it correct wrong answers}
in the corrupted training set.
The corresponding phase space of vectors is large (1746 students with a
genuine error level, i.e.~with respect to the original rule, reduced
from $E_0=5$ to $E_0=4$ in the example of section 5.1).
 \item On the other hand, the steep shore on the right means that the
phase space corresponding to detrimental changes of answers, which are
correct in the training set, should be significantly smaller (1146 students
with an enhanced error level, $E_0=5 \to E_0=6$).
\end{itemize}
 So, corresponding to the ''water dynamics'' scenario of Fig.~17, if the
noise level is enhanced, the additional accessible phase space is
dominantly situated on the left. The corresponding flat bank corresponds
to just those solutions which correct errors instead of adding new ones.
Since the former are as well those having (intuitively) an increased
overlap with the teacher, 'training with noise' can be useful.

But what happens, if $T_s$ is increased too much, corresponding to
flooding the lake~? First of all, we have to notice that the 'optimal' solution
is placed somewhere in the neighbourhood of the deepest point towards the 
flat shore.
So flooding means that many solutions far from the optimum are included,
and the performance of a random choice (corresponding to the Gibbs algorithm) 
should decrease.
Nevertheless, looking at the {\it center of mass} (corresponding to the
Bayes algorithm), this quantity should be less sensitive to the deluge,
especially if the shapes of the shores become more similar in the remote
areas of the lake. So
the Bayes algorithm should be less affected by a too high student temperature.
This lower sensitivity of the Bayes algorithm to the negative effect of a
very high noise, as contrasted to the case of Gibbsian learning, can be
nicely observed in Fig. 7.

\section{Conclusions}

We have studied the influence of  input- or  output noise on the
existence of {\it phase trans\-itions} in the generalization
behaviour of two-layer neural networks with non-overlapping receptive
fields. Generally we find for Gibbs learning as a function of the
reduced size $\al
:=p/N$ of the training set that the 'Aha-effect' phase transitions, where
the system  performs a simple guess for $\al < \al_c$ and only
generalizes if $\al $ exceeds a critical value, {\it persist} in the presence of
noise. However, the critical parameters scale with the strength of the
noise, see e.g.\ Figs.\ 3 and 4.
 In particular, the order-index $n$ of the system, which characterizes
the behaviour at the transition, is unchanged: For $n=1$ (e.g.\ the
committee machine) the system starts to
generalize already with arbitrarily small $\al$, while for $n=2$ (e.g.\ the
$K=2$-parity machine) there is a
continuous 'Aha-effect' transition (i.e.\ a 2nd order phase transition),
whereas for
$n\ge 3$ (e.g.\ the parity machine with $K\ge 3$, which has $n=K$)  a discontinuous
'Aha-effect' transition happens. Only the so-called
'interim phase transitions', which appear in some cases for $n=1$, e.g.\ the
K4\_mcode2 machine, get lost by input noise, but not by
output noise. (At the 'interim transitions' only the strength of the
generalization ability is enhanced from an already finite value for $\al <
\al_c$ to a larger value above $\al_c$). Therefore, the 'Aha-effect'
transitions are -- so to say -- more generic than
'interim-transitions'.

Concentrating on the simple perceptron and output noise we also studied
the problem of an {\it optimal choice} for the noise level $T_s$ of the
student machine, given that of the teacher. Looking at {\it Gibbsian
learning} there is a critical $\alpha_c$ above which a student trained
with a finite noise level has a better performance than a 'perfect
student', i.e.~a student who has learnt the (partially corrupted)
training set without errors.
For $\alpha\to\infty$ this optimal noise level approaches
 $\approx 0.6\,T_t$ (this is true
  for all network types considered here).  The case is
slightly different for {\it Bayesian learning}. Here the choice $T_s=T_t$ is
optimal for all $\alpha$. Nevertheless the learning curve is quite
 flat
around this optimal choice, so small deviations from the optimal noise
 have no large impact.

Concerning replica symmetry breaking for the problem considered we found
that for given $T_t$ replica symmetry is typically conserved down to
the optimal values of $T_s$ and somewhat below.
 But if $\alpha$ is larger than a critical
 noise-dependent value,
 replica symmetry breaking occurs for low noise-temperatures.
Due to this fact, the sub-optimal overlap $r$ calculated
 in a replica symmetric theory for this region of small $T_s$ is corrected
to a higher value, which in a RSB1 calculation almost reaches the optimal
number. This means that the 'gain' achieved by training with noise is less
pronounced than predicted in e.g.\ the RS theory of \cite{l:gyoergyi}.
Replica symmetry breaking of higher order may lead to additional (slight) 
corrections in the region,
% between $T_s=0$ and somewhat below the 'optimal' $T_s$,
 where replica symmetry is broken.
Nevertheless the effect of an optimal finite student noise rate $T_s$ in some
cases was
shown to be not artificial since the values of $\al$
resp.\ $T_s$, below resp.\ above which replica symmetry remains
 preserved, and also the optimal value of $T_s$ in this RS region, as
  calculated in this
paper, will remain unchanged.

Finally we took a quick look on mechanisms allowing improved learning
for imperfect students: We showed that error correction is possible,
 due to a sort of energy-entropy
competition, leading to an increased overlap
compared to the minimal-error solutions.

\subsection*{Acknowledgements}
The authors would like to thank G.\ Dirscherl, F.\ Gerl and M.\ Probst
for stimulating discussions. BS thanks the Leverhulme Trust for support
 (F/250/K). Finally we would like to thank one of the referees, who was so
kind to send us a recent preprint of T.~Uezu,
\cite{l:Uezu}, in which the problem of the optimal noise level of the
student is treated for single-layer perceptrons with different types of
noise in an RS and RSB1 approach. Uezu's results strongly overlap with
parts of our chapter 4.3.

\newpage
\centerline{\bf{Figure captions}}

{\bf Figure 1.} The architecture of the class of networks considered:
There are $N$ binary input units separated into $K$ different groups
leading each to a 'hidden unit'. The binary outputs of the 'hidden
units' are fed into a final Boolean output function. Only the weights
$\wb$ from the inputs to the hidden units can be modified by learning
processes. Two kinds of noise will be considered below: 'input noise'
and 'output noise'.

{\bf Figure 2.} The scaling factor $r(\beta)$ of Eqn.\ (\ref{eq63}), which
applies to the case $q\to 1$ of high loading, and the 'intuitive scaling
factor' $r^{\rm\scriptsize{it}}=(1-2p_f)^2$, which applies to the limit $q\to
0$, are presented as a function of
the 'corrupted fraction' 2$p_f=2e^{-\beta}/(1+e^{-\beta})$ of the training
set with respect to the output.
$\beta:=\beta_s=\beta_t=1/T_s=1/T_t$ characterizes the output noise-levels
of the student and the teacher machine.

{\bf Figure 3.} Data collapsing for output noise: On the left-hand-side, the
overlap $q(\al)$ of the couplings of the student and teacher machines
is presented as a function of the loading
 $\al =p/N$, where $p$ is the number of examples of the training
set, for 4 different values of the common output noise-level $T=1/\beta$ of
the student and the teacher ($T=0$; $0.5$; $1.0$ and $1.5$), whereas on the
right-hand-side the  results (for $T=1.0$, $1.5$, $2.0$ and $5.0$) are
presented as a function of  $\al\beta^2$. Note that for the
parity machine with $K=2$ and $3$, respectively, one has an 'Aha-Effect'
phase-transition of 2nd order (n=2) and 1st order (n=3) respectively,
whereas for the committee machine and the K4\_mcode2 machine, where the
order index $n=1$, the machine generalizes right from the beginning. For
the committee machine, there is no phase transition at all, whereas for the
K4\_mcode2 machine, there is an 'interim transition' around
$\al\beta^2\approx 16$, a situation, which is also compatible with $n=1$,
see \cite{l:scho1}. 

{\bf Figure 4.} For the $K=2$ and $K=3$ parity machines, and for the
$K=3$ committee, the prefactor $\tilde c$ of the $\tilde c/\sqrt{\al}$
-asymptotics of the error-probability $\epsilon(\alpha )$
is plotted against $\gamma$ and $p_f$, respectively, for the
case of input noise; $\gamma\equiv \beta^{-1}$ is the common
noise-level of teacher and student machines,
and $p_f$ is defined in (33) and (34). In the lower plot, the curves for
the $K=3$ parity and committee machines overlap to the accuracy of the
drawing.

{\bf Figure 5.} Data collapsing for input noise: For the $K=2$ parity
machine and for the K4\_mcode2 machine, with common noise
temperatures $\gamma=1/\beta$ of the teacher and student machine, ranging
from $\gamma=0$ via 1, 1.5, 2 to 5, the overlap $q$ between teacher and
student couplings is plotted against $\al$ and $\al/\zeta^n$, respectively,
where $n$ is the order-index of the system and $\zeta :=(1+\gamma)^{-1}$.
For the parity machine, the data collapsing is almost perfect, whereas for
the second machine it applies only to the limits of small and high values of
$\al/\zeta$. Note that here, in contrast to Fig.\ 3, the 'interim transition'
of the second machine is destroyed by the input noise.

{\bf Figure 6a.} Storage capacity $\al_c$ of a deterministic student
perceptron (i.e.\ $K=1$, $T_s=0$) in the presence of a noisy training set
($T_t\ne 0$): A fraction
$p_f$ of the binary answers of the teacher perceptron are misclassified.
Both, input- and output noise are considered.

{\bf Figure 6b.} Overlap $r$ of a deterministic student perceptron
(i.e.\ $K=1$, $T_s=0$) with the teacher vector in the presence of a
noisy training set with output noise strength $T_t=1$, plotted as a
function of the reduced size $\al :=p/N$ of the training set. The solid
line is for Maximal Stability Learning, i.e.\ the AdaTron algorithm,
\cite{l:Anlauf}, while the dashed line is for Gibbs learning.  The two
overfitting effects appearing here are explained in the text.

{\bf Figure 7.} For the perceptron ($K=1$) and the
case  $T_t=1$ of the teacher's output noise-level
and the three cases of $\al=1$, $=2$, and $=5$, the overlaps $q$ and $r$
for Gibbs learning, and
$r_{\rm\scriptsize{cp}}$ for the central-point network, i.e.~Bayesian
learning, as explained
in the text, are plotted over the student machine's output noise-level
$T_s$. Note that from $r$ an {\it optimal} noise temperature
$T_s=T_{opt}$ can be defined.

{\bf Figure 8.} Here the optimal student machine's output noise temperature
$T_{opt}$, as determined in Fig.~7, is presented over $\al$ for $T_t=1$
(Gibbs learning, $K=1$). The dashed line is the asymptotic limit for
$\al\to\infty$.

{\bf Figure 9.} For the perceptron ($K=1$) with Gibbs learning, for
various values of $\al$ and fixed teacher machine's output noise
temperature $T_t=1$, the training error $\eps_{tr}$ of Eqn.\
(\ref{eq81}) is plotted over the output noise temperature $T_s$ of the
student machine.

{\bf Figure 10.} For Gibbs learning with $\al=5$, $T_t=1$ and $K=1$ the
order parameters $r^{\rm\scriptsize{RS}}$ and $q$ (for replica
symmetry), and $r^{\rm\scriptsize{RSB}}$, $q_0 $ and $q_1$ (in 1-step
replica symmetry breaking) are presented over the student perceptron's
output noise temperature $T_s$. For values of $T_s$ which are slightly
smaller than the optimal value $T_s\approx 0.35$, replica symmetry is
broken and the overlap $r^{\rm\scriptsize{RSB}}$ is somewhat enhanced
with respect to the RS case, but still smaller than the optimum.

{\bf Figure 11.} For $T_t=1$ and $K=1$, the optimal overlap
$r(T_s=T_{opt}(\al );\al)$ is
presented. For comparison, also the overlap obtained for $T_s=0.1$, where
replica symmetry is broken, is plotted, both in RS approximation and in
RSB1, where the result is only slightly sub-optimal.

{\bf Figure 12.} This figure suggests an analogy, making plausible that
replica symmetry is broken for small student machine's noise-level(left
scenario), but is {\it restored} beyond a critical value of $T_s$, for
fixed teacher machine's noise-level$T_t$ (right scenario).

{\bf Figure 13.} For a general Boolean function $B$ and the case of
output noise, the ratio of the prefactors
$c_0/c_0^{\rm\scriptsize{opt}}$ for the asymptotic behavior
$\eps(\al\to\infty)\to c_0/\al$ is plotted against $T_s/T_t$
 for the values of $T_t=0.2$, $0.5$,
$1.0$ and $2.0$. Note that the optimal value $T_s \approx 0.6 T_t$ is
more or less universal, and the behaviour in the vicinity of this point
is rather flat.

{\bf Figure 14.} Comparison of the overlap $r(\al=2)$ of our theory for
$K=1$ as a function of $T_s$ for $K=1$ and $T_t=0.91$ (dashed line,
calculated as in Fig.\ 7) with results from a simulation of the states
of a small system as described in the text (solid line). In both cases
there is an optimal output noise strength $T_s$. For more details see
the text.

{\bf Figure 15.} For a teacher output noise-level of $T_t=1$ and
$\alpha=2$ the curves for $r$, $\epsilon_0$ and $\epsilon_{tr}$ are
shown for $K=1$ as a function of the noise-level $T_s$ of the student
machine. The increase in $r$ is related to a decrease of
$\epsilon_0$ showing the error correcting behaviour of the non-perfect
student.

{\bf Figure 16.} Phase space analysis according to the formalism of
\cite{l:monasson,l:zecchina} of a perceptron (i.e.\ $K=1$)
 with $T_t=1$ and $\al
= 5$, i.e.\ the overlap $r(k)$ has been presented as a function of the
typical volume measure (-k) as explained in the text. The student
output noise temperatures of the different curves correspond to $T_s$ =1.0,
0.8, 0.6, 0.4, 0.3, 0.2, 0.16, 0.14, 0.12 and 0.1, from the left.
 The asterisk denotes the
'typical result' as obtained according to the usual RS calculation.

{\bf Figure 17.} This sketch shall make plausible, why training with noise
can lead to an enhancement of the overlap of the couplings of the student
and the teacher machine's couplings. For details see the text.
%%%%%%%%%%%%%%%%%%%%%%%%%%%%%%%%%%%%%%%%%%%%%%%%%%%%%%%%%%%%%%%%%%%%%%%%
%\end{document} %If You don' want the drawings, delete the % in this line
\newpage 
\input epsf
\epsfxsize=15cm
\epsfbox{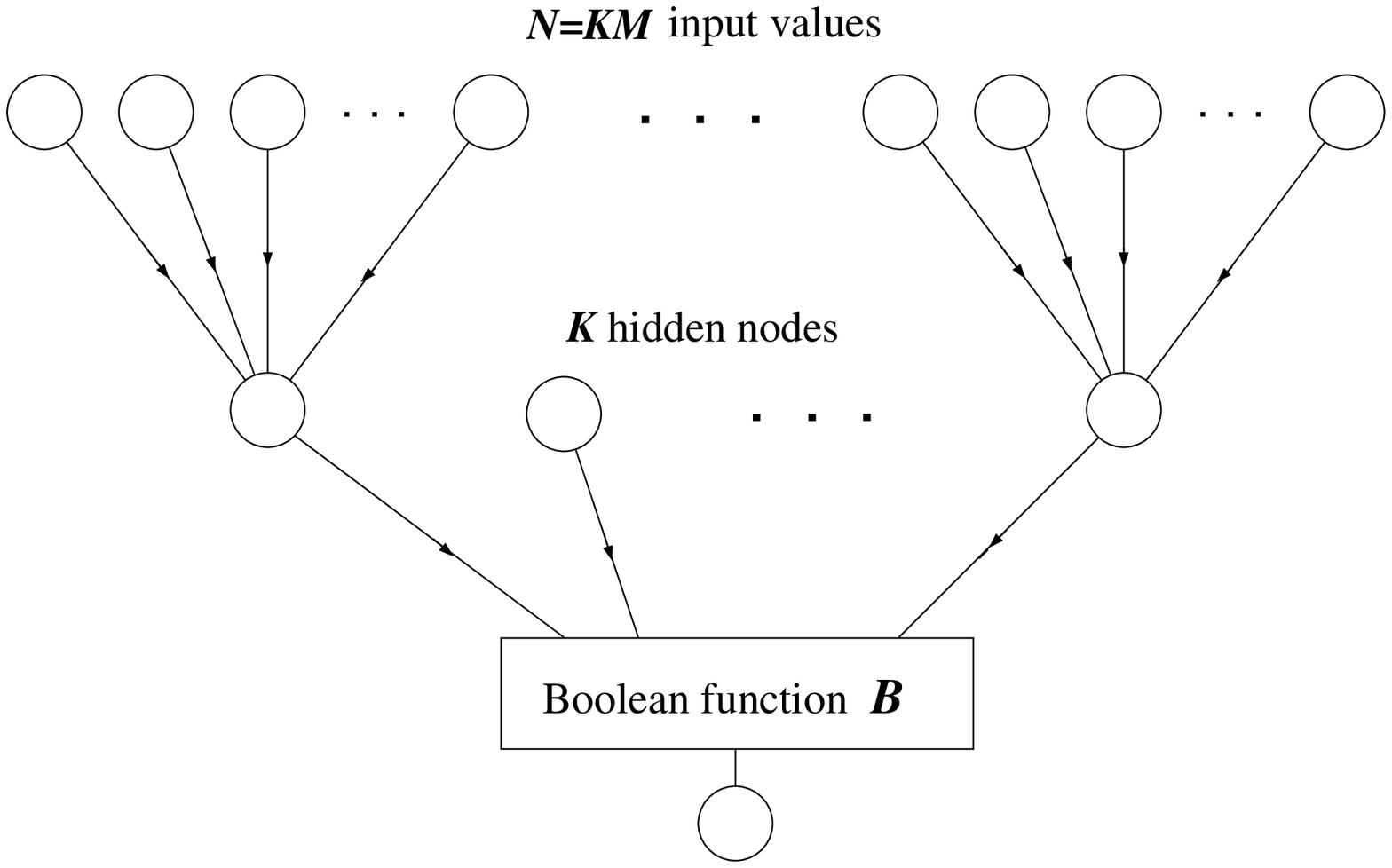}
{\centerline{\underbar{Fig.~1}}}
\vglue 0 truecm
%%%%%%%%%%%%%%%%%%%%%%%%%%%%%%%%%%%%%%%%%%%%%%%%%%%%%%%%%%%%%%%%%
\epsfxsize=16cm
\epsfbox{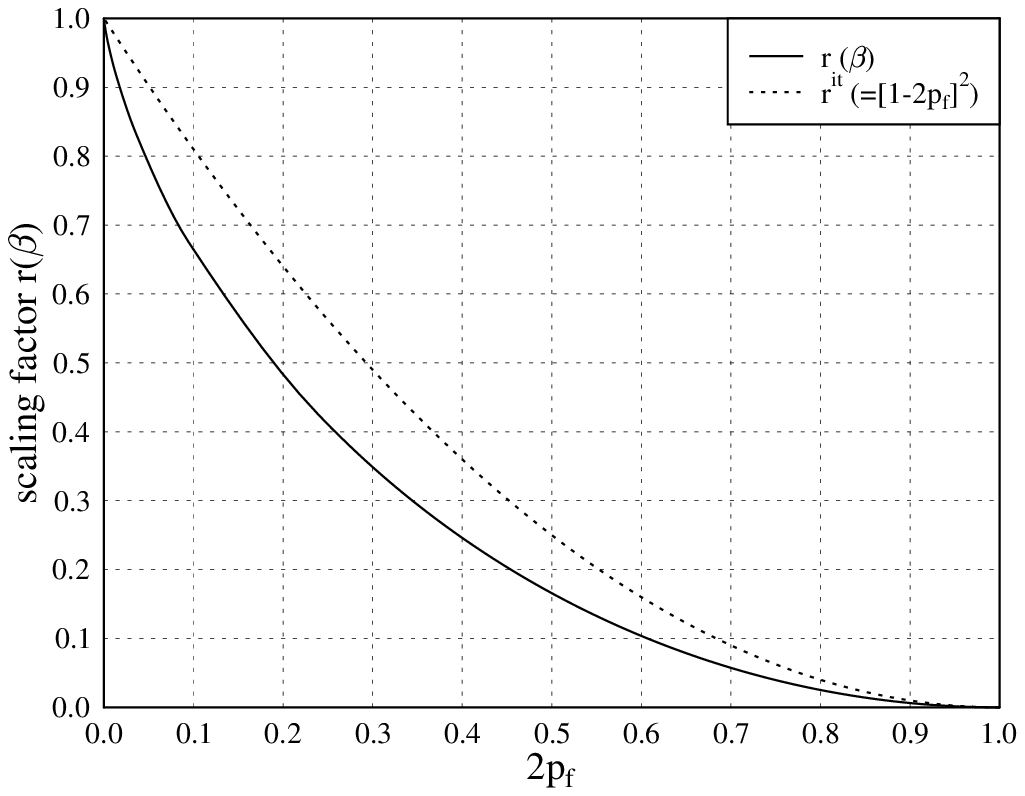}
\centerline{\underbar{Fig.~2}} 
\newpage 
%%%%%%%%%%%%%%%%%%%%%%%%%%%%%%%%%%%%%%%%%%%%%%%%%%%%%%%%%%%%%%%%%
\epsfxsize=14.5cm
\epsfbox{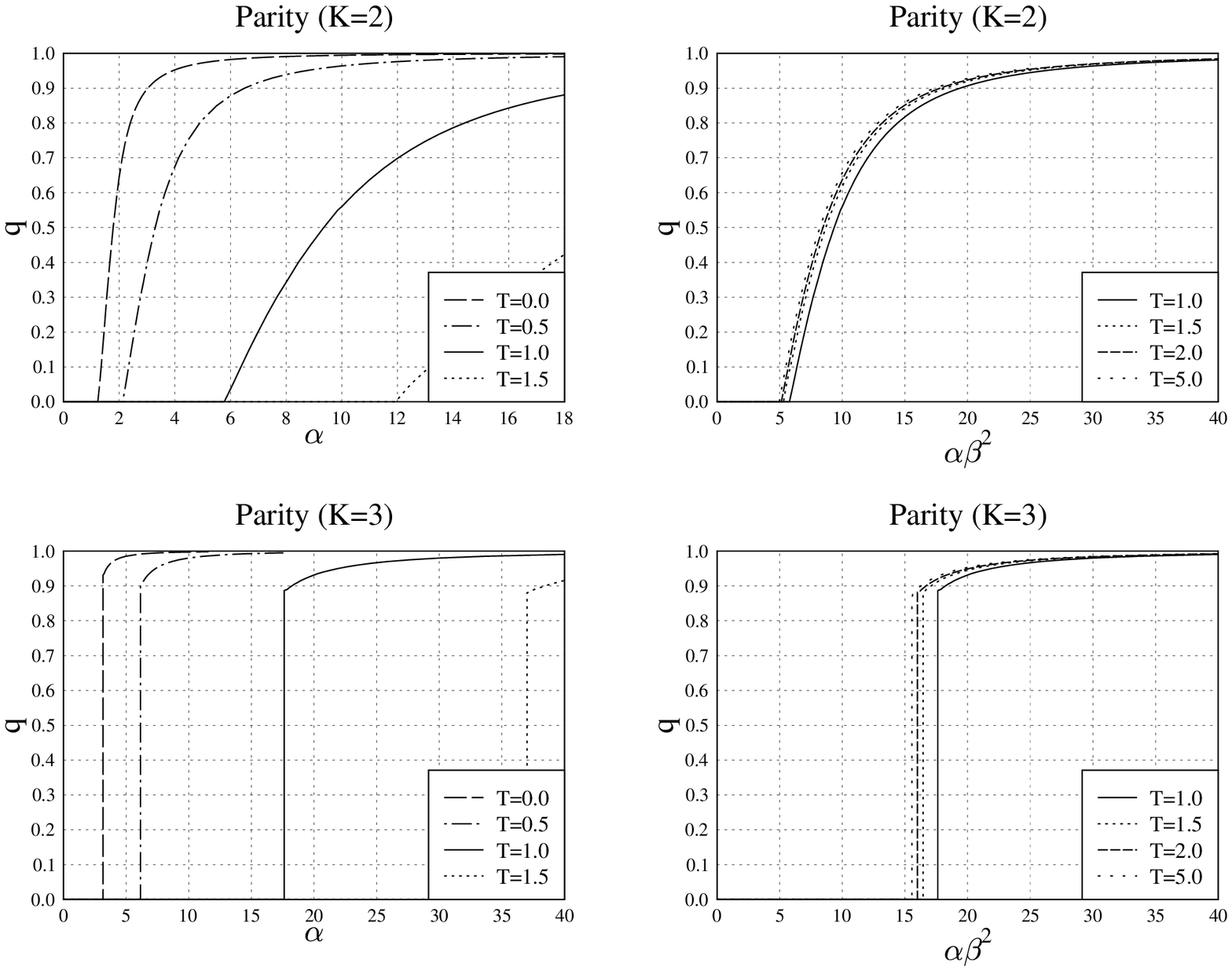}
\vglue 0 truecm
\epsfxsize=14.5cm 
\epsfbox{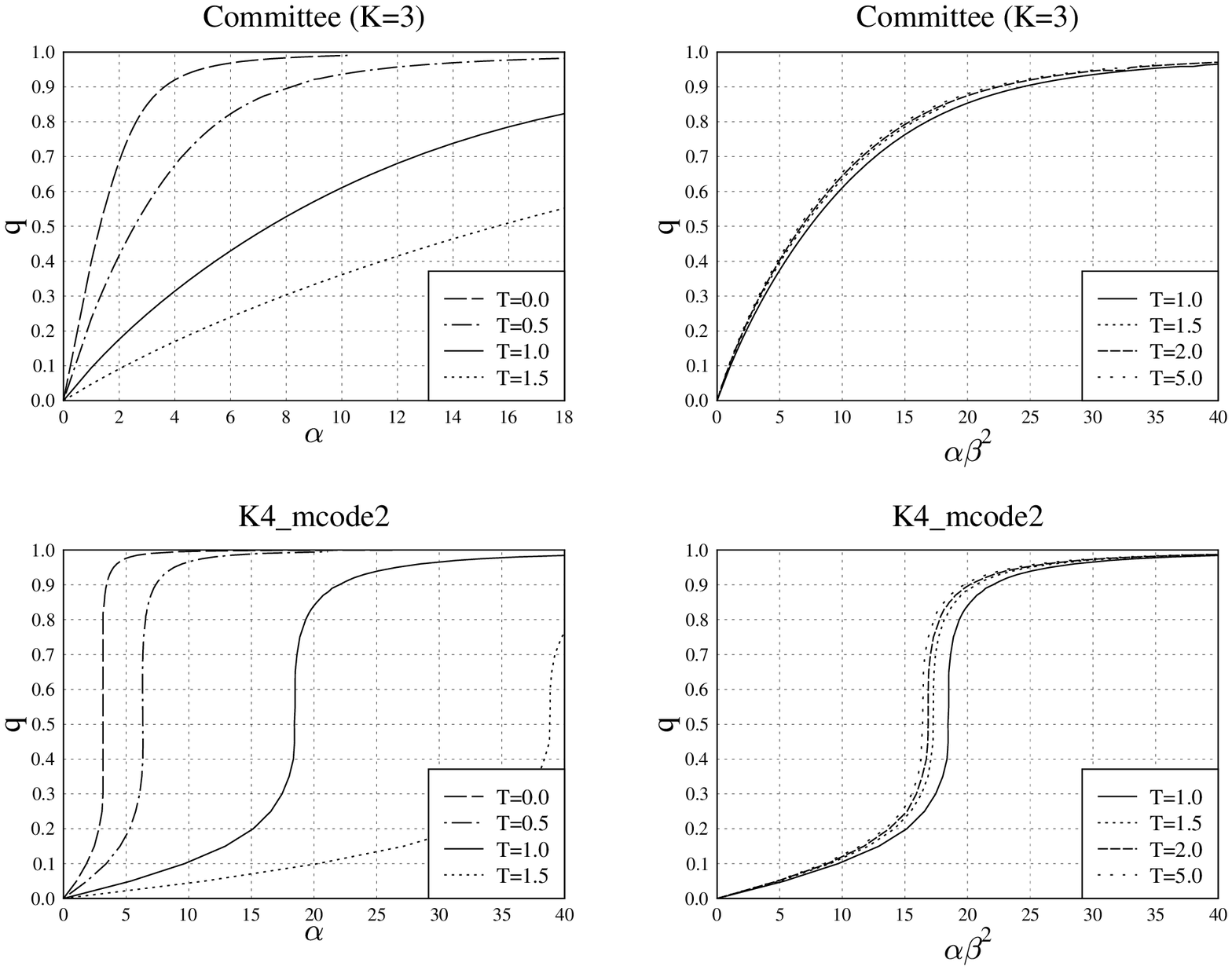}
\centerline{\underbar{Fig.~3}} 
%%%%%%%%%%%%%%%%%%%%%%%%%%%%%%%%%%%%%%%%%%%%%%%%%%%%%%%%%%%%%%%%%
\epsfxsize=16cm
\epsfbox{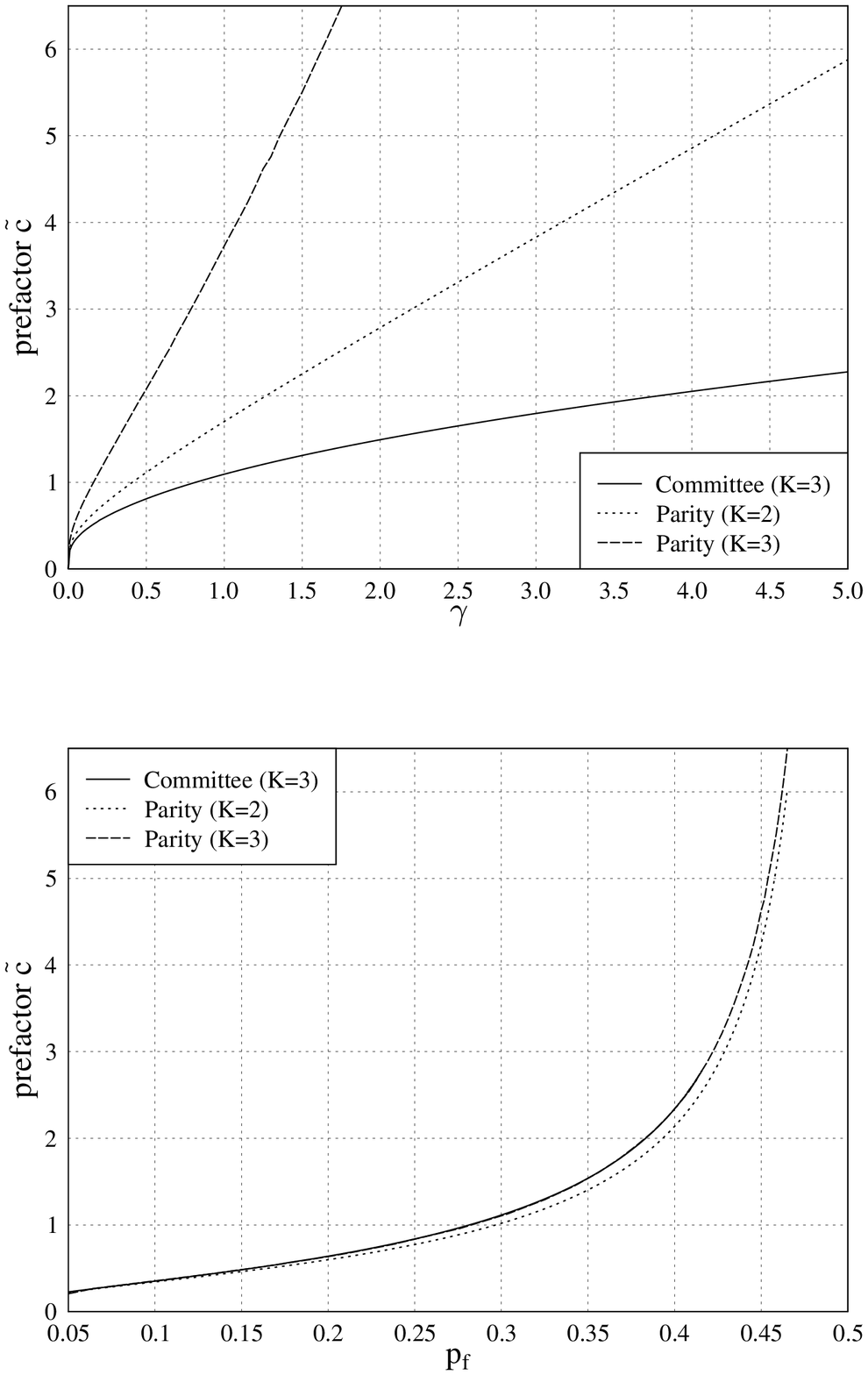}
\centerline{\underbar{Fig.~4}}
%%%%%%%%%%%%%%%%%%%%%%%%%%%%%%%%%%%%%%%%%%%%%%%%%%%%%%%%%%%%%%%%%
\epsfxsize=16cm
\epsfbox{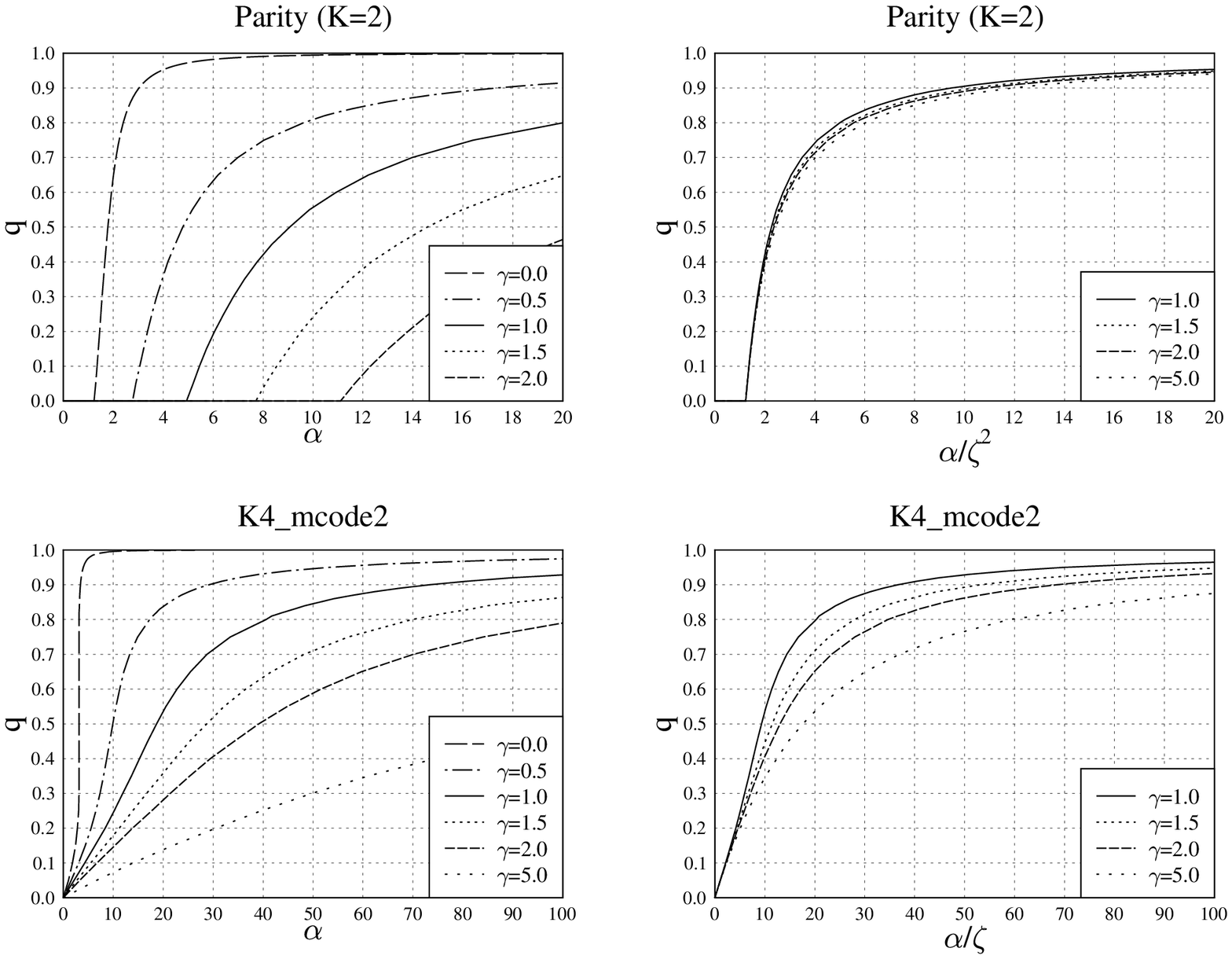}
\centerline{\underbar{Fig.~5}}
%%%%%%%%%%%%%%%%%%%%%%%%%%%%%%%%%%%%%%%%%%%%%%%%%%%%%%%%%%%%%%%%%
\epsfxsize=14.5cm
\epsfbox{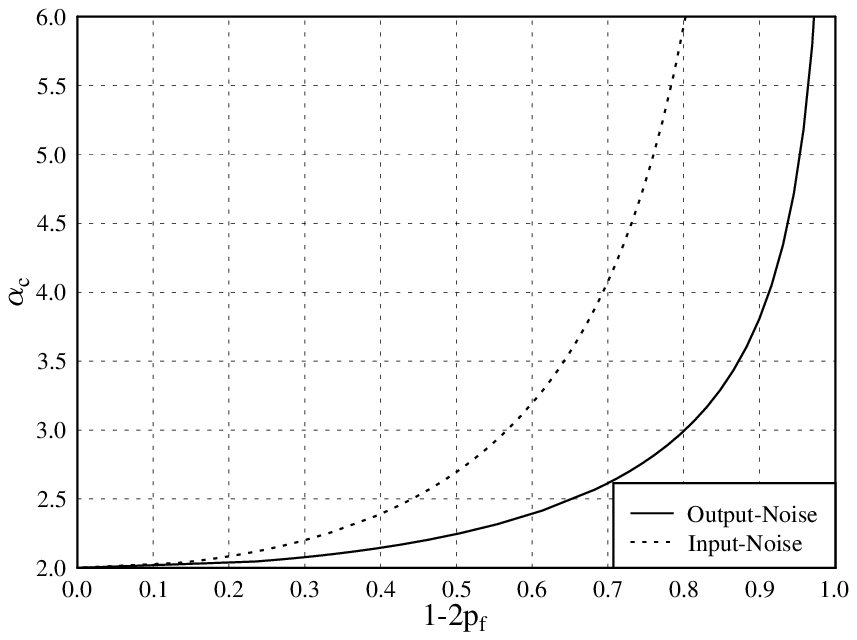}
\centerline{\underbar{Fig.~6a}} 
\epsfxsize=14.5cm
\epsfbox{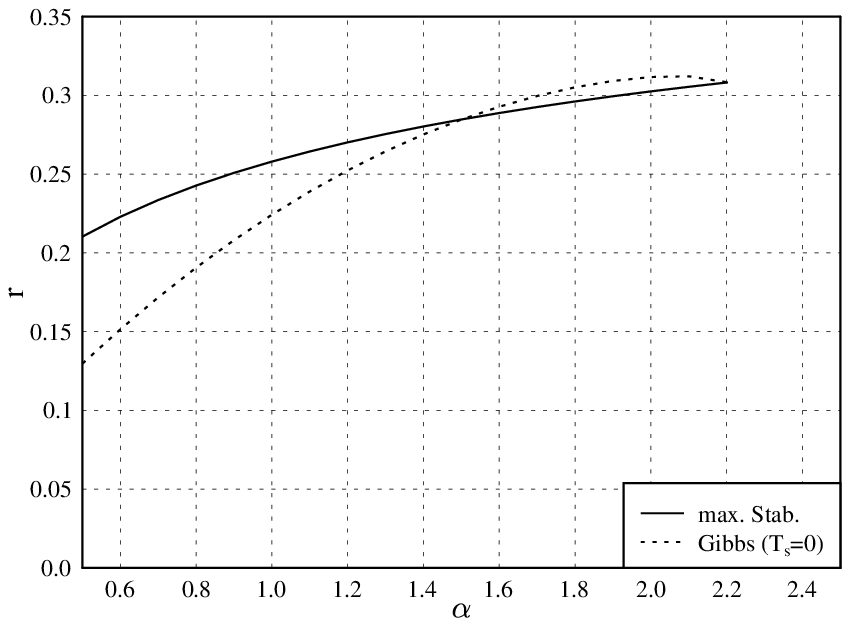}
\centerline{\underbar{Fig.~6b}} 
%%%%%%%%%%%%%%%%%%%%%%%%%%%%%%%%%%%%%%%%%%%%%%%%%%%%%%%%%%%%%%%%%
\epsfxsize=10.0cm
\epsfbox{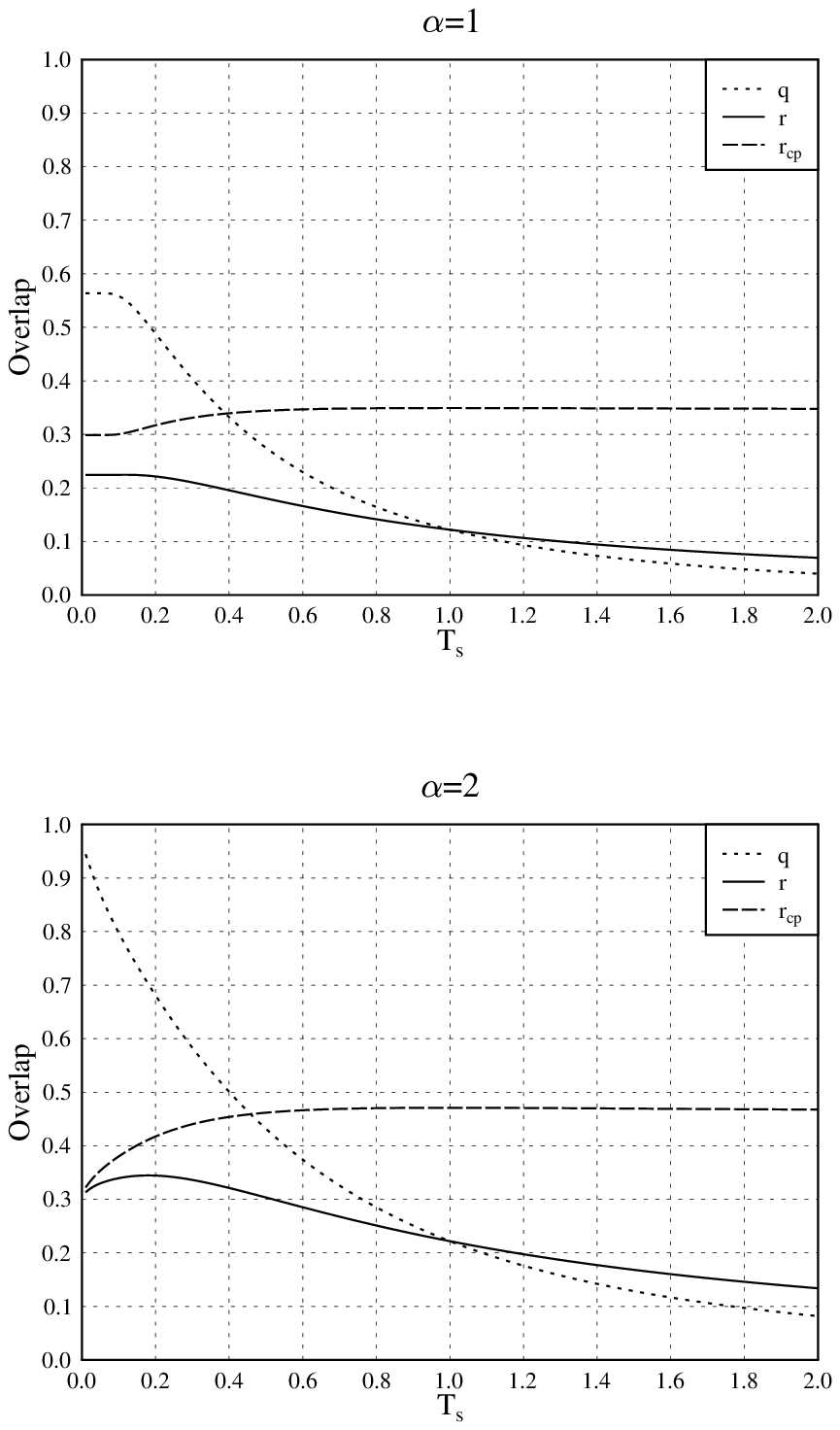}
\vglue 0 truecm
\epsfxsize=10.0cm 
\epsfbox{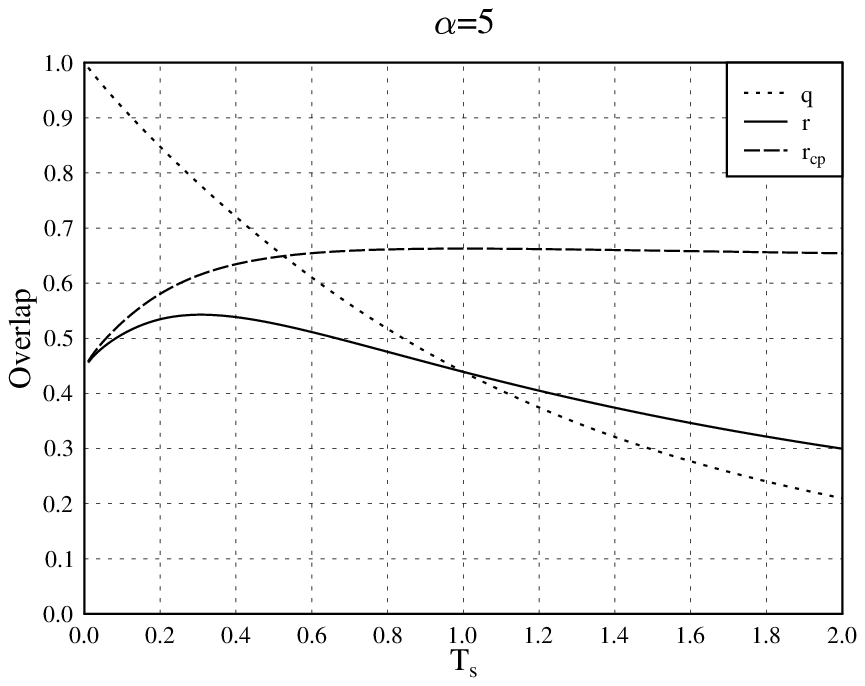}
\centerline{\underbar{Fig.~7}} 
%%%%%%%%%%%%%%%%%%%%%%%%%%%%%%%%%%%%%%%%%%%%%%%%%%%%%%%%%%%%%%%%%
\epsfxsize=14.5cm
\epsfbox{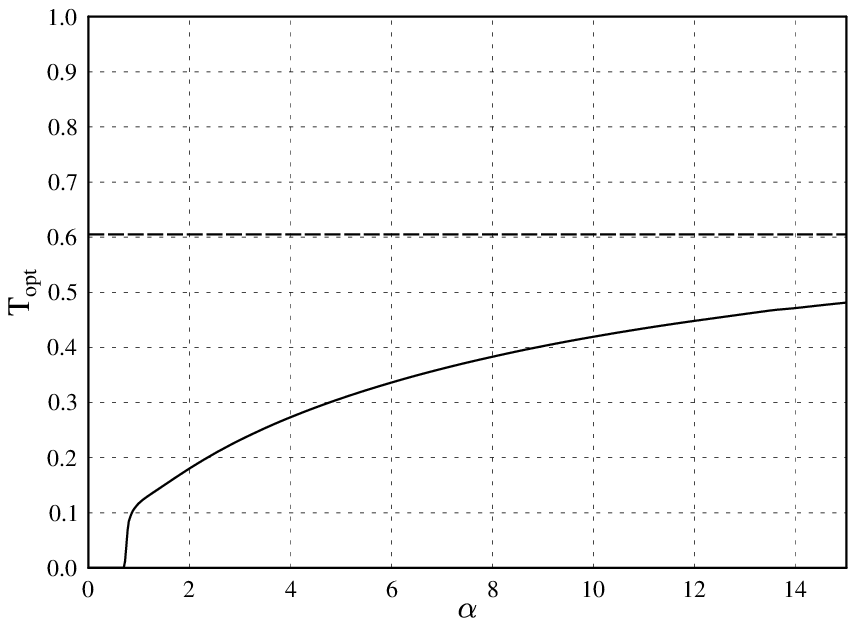}
\centerline{\underbar{Fig.~8}}
\vglue 0 truecm 
%%%%%%%%%%%%%%%%%%%%%%%%%%%%%%%%%%%%%%%%%%%%%%%%%%%%%%%%%%%%%%%%%
\epsfxsize=14.5cm
\epsfbox{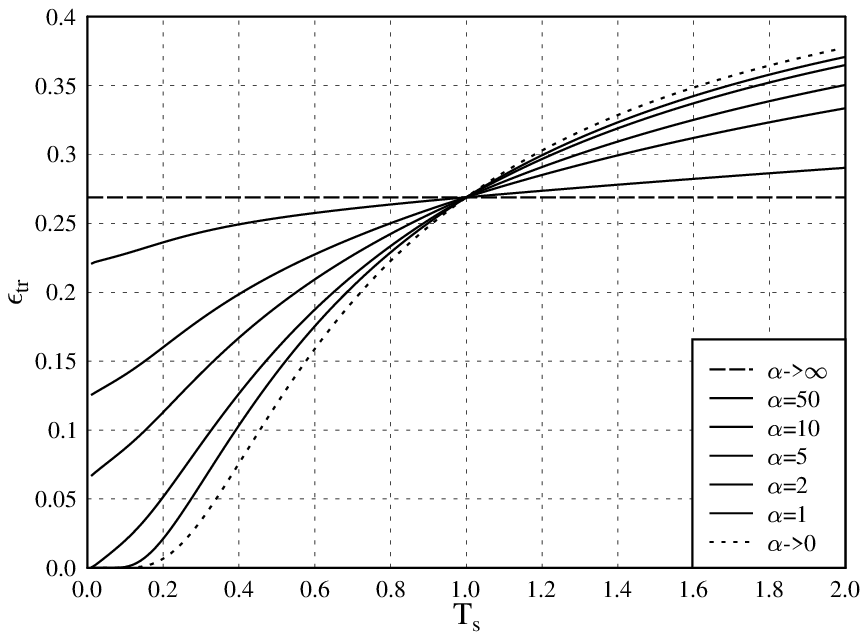}
\centerline{\underbar{Fig.~9}}
%%%%%%%%%%%%%%%%%%%%%%%%%%%%%%%%%%%%%%%%%%%%%%%%%%%%%%%%%%%%%%%%%
\epsfxsize=17.0cm
\epsfbox{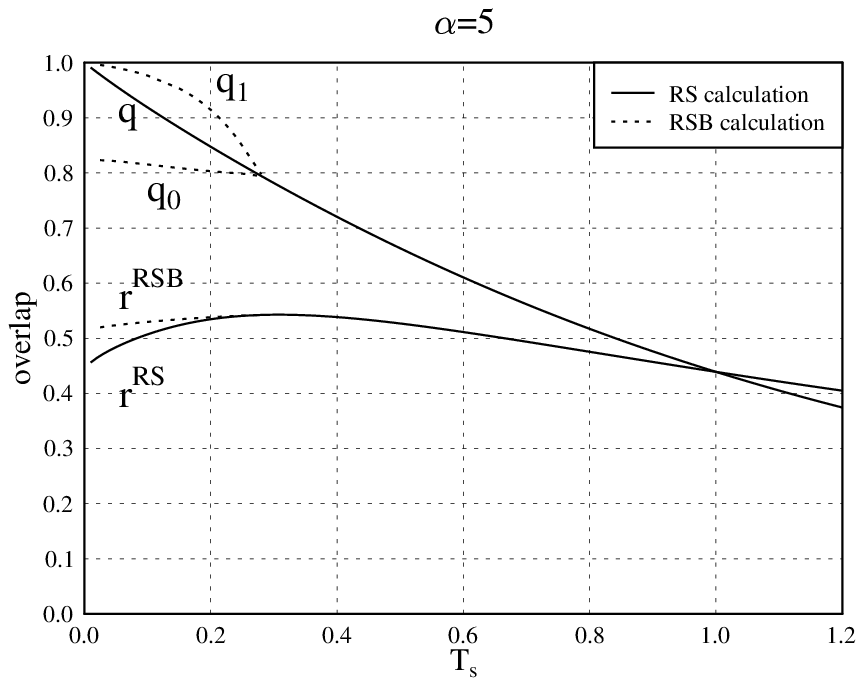}
\centerline{\underbar{Fig.~10}} 
%%%%%%%%%%%%%%%%%%%%%%%%%%%%%%%%%%%%%%%%%%%%%%%%%%%%%%%%%%%%%%%%%
\epsfxsize=15.0cm
\epsfbox{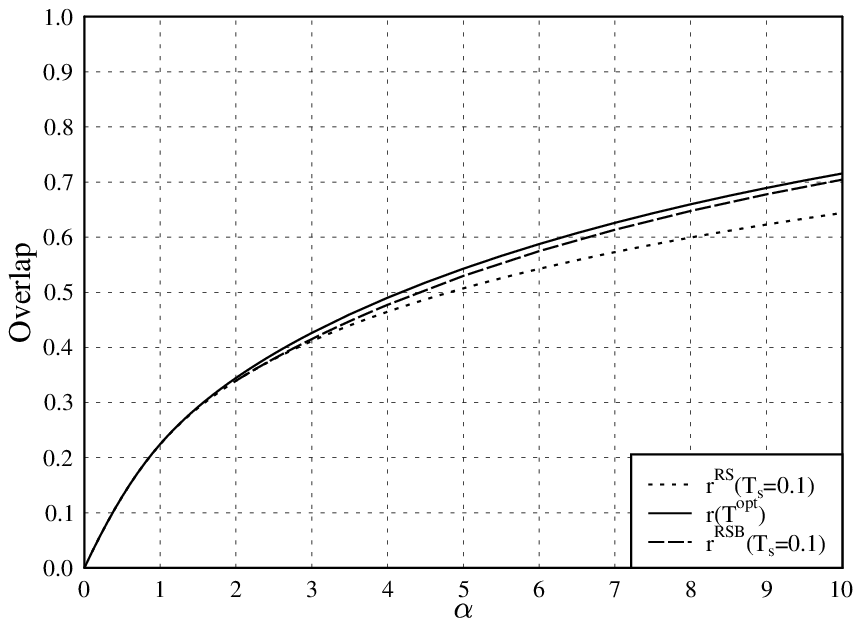}
\centerline{\underbar{Fig.~11}}  
%%%%%%%%%%%%%%%%%%%%%%%%%%%%%%%%%%%%%%%%%%%%%%%%%%%%%%%%%%%%%%%%%
\vglue 1 truecm
\epsfxsize=16cm
\epsfbox{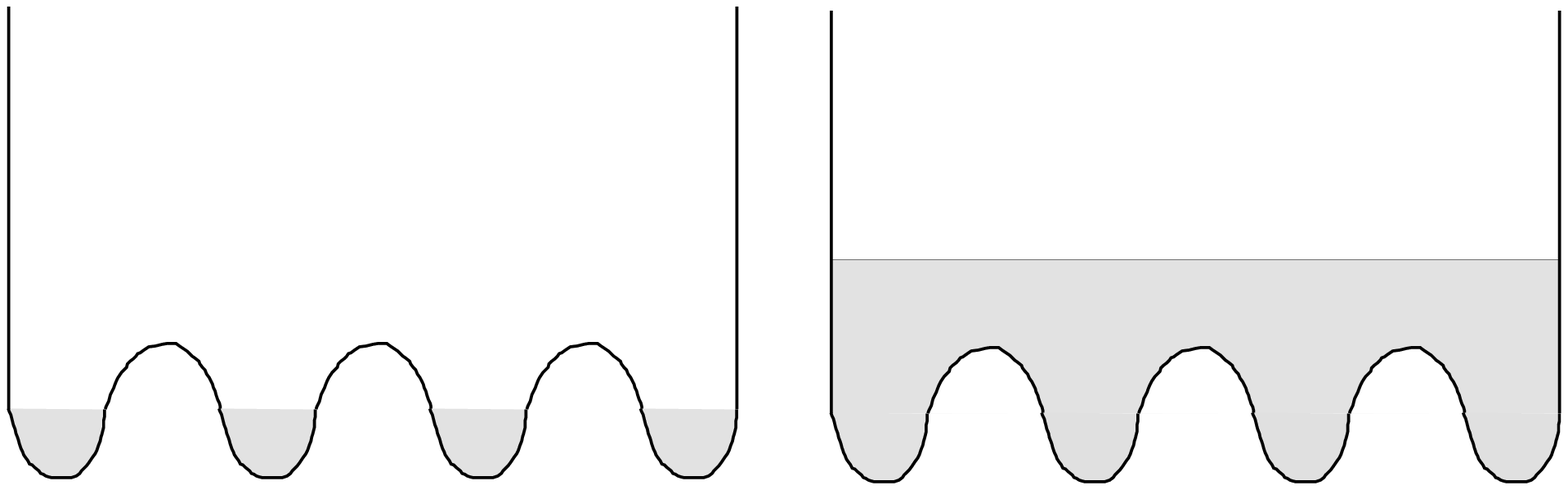}
\vglue 1 truecm
\centerline{\underbar{Fig.~12}}
%%%%%%%%%%%%%%%%%%%%%%%%%%%%%%%%%%%%%%%%%%%%%%%%%%%%%%%%%%%%%%%%%
\epsfxsize=15.0cm
\epsfbox{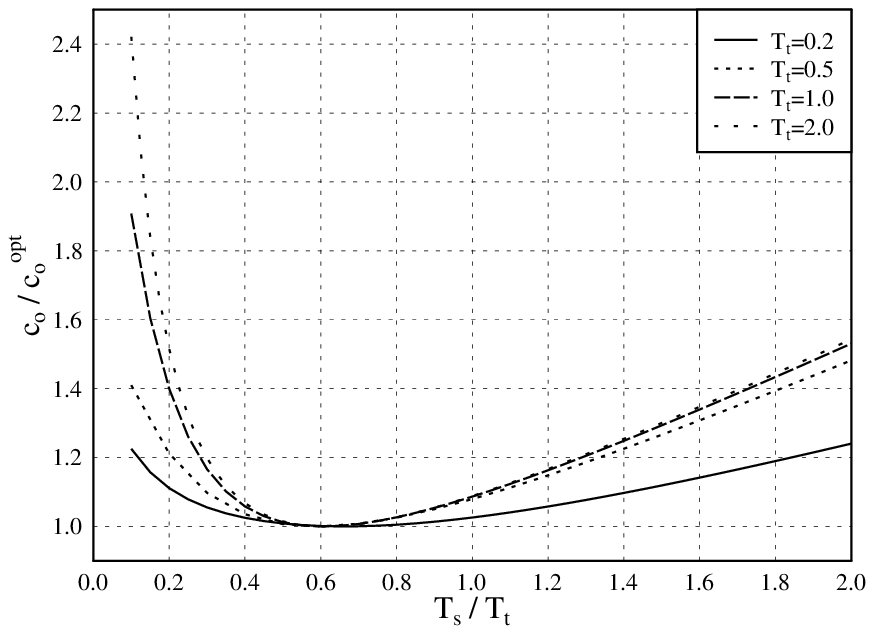} 
\centerline{\underbar{Fig.~13}}
%%%%%%%%%%%%%%%%%%%%%%%%%%%%%%%%%%%%%%%%%%%%%%%%%%%%%%%%%%%%%%%%%
\epsfxsize=15.0cm
\epsfbox{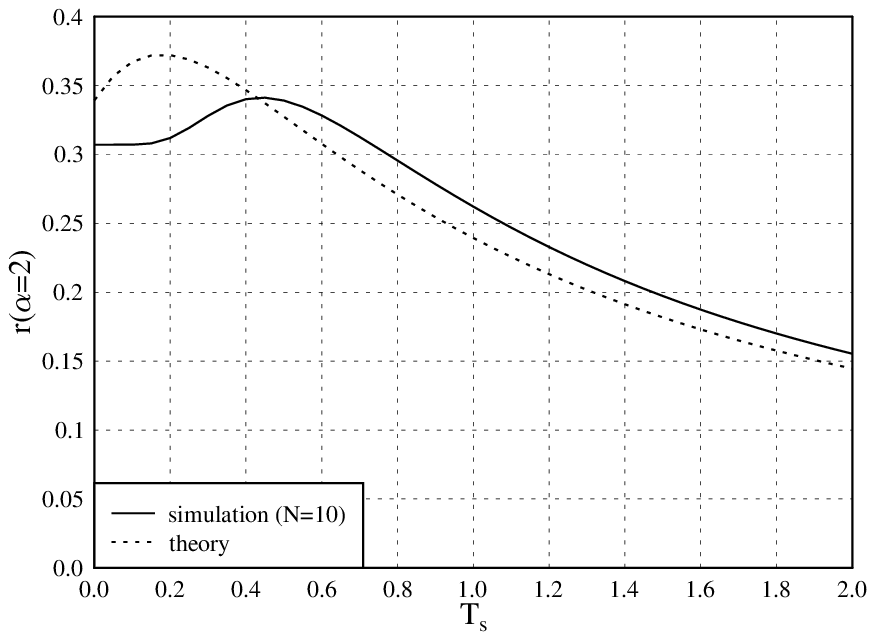}
\centerline{\underbar{Fig.~14}}
%%%%%%%%%%%%%%%%%%%%%%%%%%%%%%%%%%%%%%%%%%%%%%%%%%%%%%%%%%%%%%%%%
\epsfxsize=15.0cm
\epsfbox{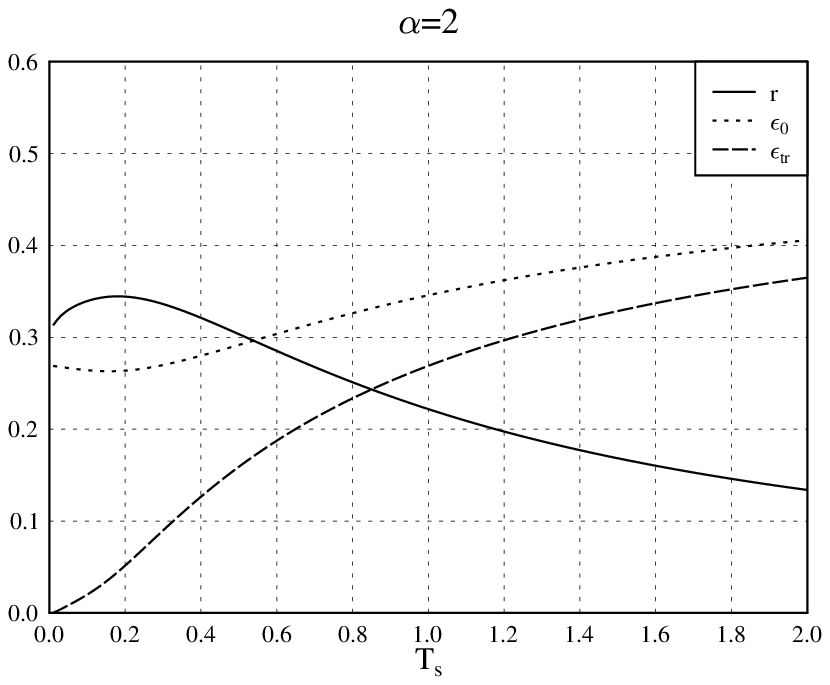} 
\centerline{\underbar{Fig.~15}}
%%%%%%%%%%%%%%%%%%%%%%%%%%%%%%%%%%%%%%%%%%%%%%%%%%%%%%%%%%%%%%%%%
\epsfxsize=16cm
\epsfbox{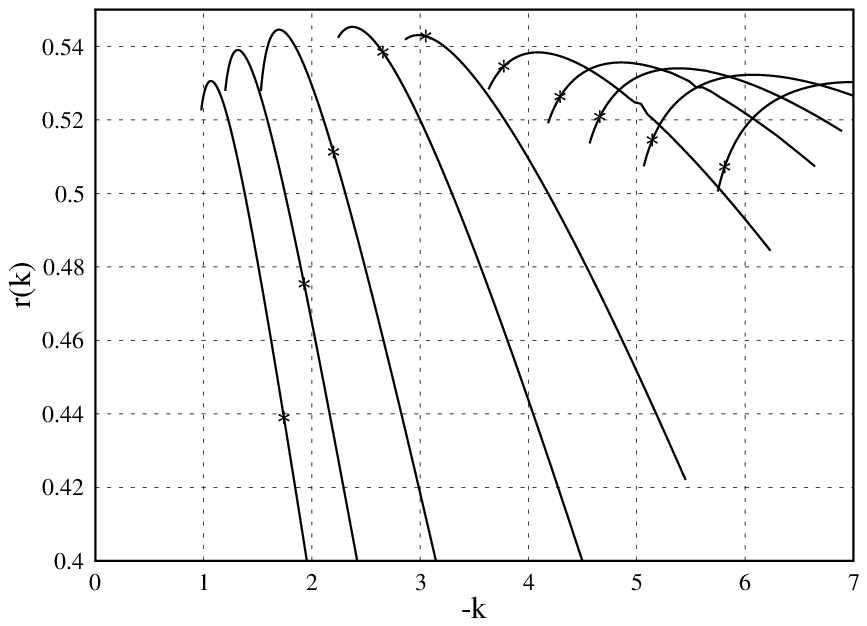}
\centerline{\underbar{Fig.~16}}
%%%%%%%%%%%%%%%%%%%%%%%%%%%%%%%%%%%%%%%%%%%%%%%%%%%%%%%%%%%%%%%%%
\epsfxsize=16cm
%\epsfbox{/home/krey/scb04349/diss/graph/Miscnoise/asytal.eps}
\epsfbox{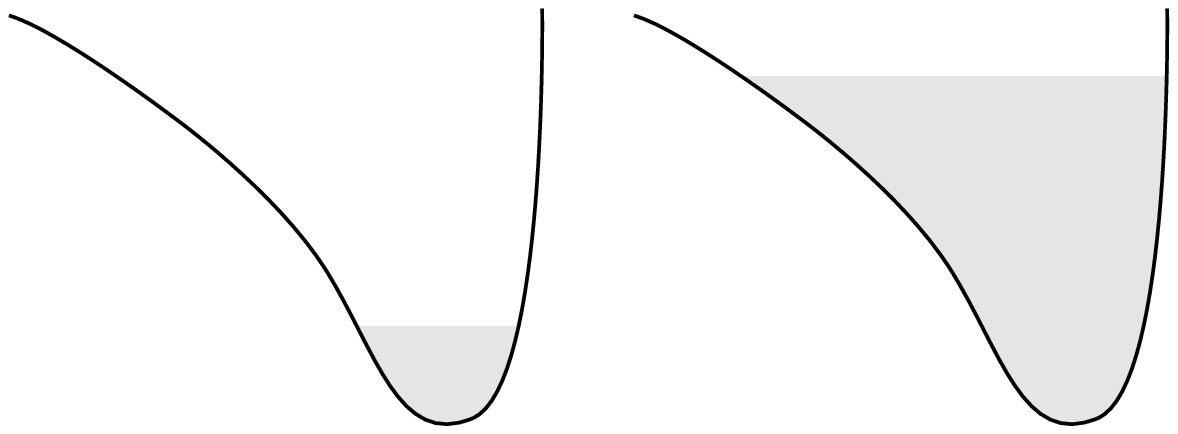}
\centerline{\underbar{Fig.~17}} 
 

\begin{thebibliography}{99}
 \bibitem{l:scho1} Schottky B 1995 Phase transitions in the
 generalization behaviour of multilayer neural networks {\it J. Phys. A:
 Math. Gen.} {\bf 21} 4515-31

 \bibitem{l:watkin} Watkin T H L, Rau A, and Biehl M 1993 The statistical
mechanics of learning a rule {\it Rev. Mod. Phys.} {\bf 65} 499-556

 \bibitem{l:opperkinzel} Opper M and Kinzel W
 1995 Statistical mechanics of generalization; in: {\it Physics of neural
 networks} ed J L van Hemmen, E Domany and K Schulten (Berlin:
 Springer), Vol.\ III 

\bibitem{l:HanselMato} Hansel D, Mato G and Meunier C 1992
Memorization without Generalization in a Multilayered Neural Network,
{\it Europhys. Lett.} {\bf 20} 471-476"


\bibitem{l:Schwarze} Schwarze H 1993 
Learning a rule in a multilayer neural network
{\it J. Phys. A} {\bf 26} 5781-94

\bibitem{l:Opper} Opper M 1994
Learning and generalization in a two-layer neural network: The role of
the Vapnik-Chervonenkis dimension, {\it Phys.Rev.Lett.} {\bf 72} 2113-21
 
\bibitem{l:gardner} Gardner E 1987 Maximum storage
 capacity in neural networks {\it Europhys.\ Lett.\ } {\bf 4} 481-5

\bibitem{l:schottkydiss} Schottky B 1996 {Generalization behaviour of
multilayer neural networks},
 {\it PhD thesis}, University of Regensburg 1996, in
German

\bibitem{l:mezard} Mezard M, Parisi G, and Virasoro M A 1987 Spin glass
theory and beyond (Singapore: World Scientific)

\bibitem{l:Anlauf} Anlauf J K, Biehl M 1989 The AdaTron: An adaptive
perceptron algorithm  {\it Europhys. Lett.} {\bf 10} 687-92

\bibitem{l:gyoergyi} Gy{\"o}rgyi G 1990 Inference of a Rule by a Neural 
Network with Thermal Noise {\it Phys. Rev. Lett.} {\bf 64} 1957-60

\bibitem{l:poeppel} P\"oppel G., Krey U. 1987 A dynamical Learning
process for the recognition of correlated patterns in symmetric spin
glass models {\it Europh. Lett.} {\bf 4} 979-85

\bibitem{l:stroud} Gardner E., Stroud N, Wallace D J 1989 Training with
noise {\it J. Phys. A: Math. Gen.} {\bf 22} 2019-30

\bibitem{l:nehl} Opper M., Kinzel W, Kleinz J, and Nehl R 1990 On the
ability of the optimal perceptron to generalize {\it J. Phys. A: Math.
Gen.} {\bf 23} 581-86

\bibitem{l:monasson} Monasson R, O'Kane D 1994 Domains of solutions and
replica symmetry breaking in multilayer neural networks {\it Europh.
Lett.} {\bf 27} 85-90

\bibitem{l:zecchina} Monasson R, Zecchina R 1995 Weight space structure
and internal representations: A direct approach to learning and
generalization in multilayer neural networks {\it Phys. Rev. Lett.} {\bf
75} 2432-35

\bibitem{l:Uezu} Uezu T 1997 Learning from stochastic rules under finite
temperature - Optimal temperature and Asymptotic learning curve - ,
submitted to {\it J. Phys. A}


 \end{thebibliography}
\end{document}